\documentclass[reprint,superscriptaddress,amsmath,amssymb,aps,prb,citeautoscript,10pt]{revtex4-2}

\usepackage[utf8]{inputenc}% Language package
\usepackage{graphicx}% Include figure files
\usepackage{bm}% Bold math
\usepackage[colorlinks,allcolors=blue]{hyperref}% Add hypertext capabilities
\graphicspath{{./figs/}}

\usepackage{xcolor}

\usepackage{titlesec}
\titleformat{\section}{\normalfont\fontsize{12}{12}\bfseries}{}{}{}[]
\titlespacing{\section}{0em}{10pt plus 8pt minus 4pt}{3pt}
\titleformat{\subsection}[runin]{\normalfont\fontsize{10}{10}\bfseries}{}{}{}[.]
\titlespacing{\subsection}{0em}{6pt plus 4pt minus 4pt}{4.4pt}
\titlespacing{\paragraph}{0em}{6pt plus 2pt minus 2pt}{4.4pt}
\usepackage{enumitem}
\usepackage{tabularx}

\allowdisplaybreaks
\begin{document}
\setcitestyle{super}  % A nature-like citation style

\title{Emulating Quantum Dynamics with Neural Networks via Knowledge Distillation}

\author{Yu Yao}
    %\email{yaoyu@usc.edu}
    \affiliation{Department of Physics and Astronomy, University of Southern California, 920 Bloom Walk, Los Angeles, CA 90089, USA}

\author{Chao Cao}
    %\email{chaocao@usc.edu}
    \affiliation{Department of Physics and Astronomy, University of Southern California, 920 Bloom Walk, Los Angeles, CA 90089, USA}

\author{Stephan Haas}
    %\email{shaas@usc.edu}
    \affiliation{Department of Physics and Astronomy, University of Southern California, 920 Bloom Walk, Los Angeles, CA 90089, USA}

\author{Mahak Agarwal}
    \affiliation{Department of Computer Science,  University of Southern California, 941 Bloom Walk, Los Angeles, California 90089, USA}

\author{Divyam Khanna}
    \affiliation{Department of Computer Science,  University of Southern California, 941 Bloom Walk, Los Angeles, California 90089, USA}

\author{Marcin Abram}
    \email{mjarbam@usc.edu}
    \affiliation{Department of Physics and Astronomy, University of Southern California, 920 Bloom Walk, Los Angeles, CA 90089, USA}
    \affiliation{Information Sciences Institute, University of Southern California, 4676 Admiralty Way, Marina del Rey, CA 90292, USA}

%%%%%%%%%%%%%%%%%%%%%%%%%%%%%%%%%%%%%%%%%%%%%%%%%%%%%%%%%%%%%%%%%:~

\date{\today}

\begin{abstract}% Typically 150 words
    High-fidelity quantum dynamics emulators can be used to predict the time evolution of complex physical systems.
    Here, we introduce an efficient training framework for constructing
    machine learning-based emulators.
    Our approach is based on the idea of knowledge distillation and uses elements of curriculum learning. It works by constructing a set of simple, but rich-in-physics training examples (a curriculum). These examples are used by the emulator to learn the general rules describing the time evolution of a quantum system (knowledge distillation).
    The goal is not only to obtain high-quality predictions, but also to examine the process of how the emulator learns the physics of the underlying problem. This allows us to discover new facts about the physical system, detect symmetries, and measure relative importance of the contributing physical processes.
    We illustrate this approach by training an artificial neural network to predict the time evolution of quantum wave packages propagating through a potential landscape. We focus on the question of how the emulator learns the rules of quantum dynamics from the curriculum of simple training examples and to which extent it can generalize the acquired knowledge to solve more challenging cases.
\end{abstract}

\maketitle

%%%%%%%%%%%%%%%%%%%%%%%%%%%%%%%%%%%%%%%%%%%%%%%%%%%%%%%%%%%%%%%%%:~

    Neural Networks can be viewed as sophisticated pattern recognition methods, capable of constructing non-linear mappings between a specified set of input data and a set of target outputs \cite{Bahri2020}.
    Their strength comes from the fact, that even with just one hidden layer they can be trained to approximate any finite, Borel-measurable function \cite{Hornik1989}.
    However, with increasing complexity of the target function, the number of required neurons in the hidden layer becomes prohibitively large \cite{Delalleau2011}.
    Therefore, it is practical to train multi-layer networks that are much more efficient in that regard \cite{Poole2016}.
    Lower layers of such deep architectures can learn efficient representation of the input data \cite{LeCun2015}, whereas the upper layers model higher-level concepts and solve the final classification or regression task \cite{Bau2020}.
    
    Neural networks, with their variety of architectures \cite{Sengupta2020}, have already been shown  to be  effective tools in  medical \cite{Jiang2017}, business \cite{great2018ai} and scientific applications, including physics in general \cite{carleo2019machine}, as well as material  \cite{Schmidt2019} and quantum science \cite{Carrasquilla2020} in particular.
    Many standard applications of machine learning methods can be reduced to either classification or regression tasks \cite{Alzubaidi2021}. In the latter case, they  can serve as powerful interpolation tools \cite{2008.03703, Chai2020}. However, out-of-domain predictions, are typically challenging \cite{Haley1992, 2009.11848}.
    While in many applications machine learning models come as theory-agnostic tools \cite{carleo2019machine, Han2018}, there also exist families of physics-informed models \cite{Karniadakis2021} that implicitly incorporate domain knowledge about the studied system, e.g., by imposing a set of constraints that preserve conservation laws or symmetries.
    In the scientific context, machine learning can be also used to enable concept discoveries. This can be accomplished by either directly constructing a machine learning system capable of answering scientific questions \cite{Iten2020} or indirectly, e.g., by interpreting already trained models \cite{Murdoch2019, Joshi2021}.
    
    In this work, we focus on the application of machine learning to the emulation of quantum dynamics.
    The goal is to examine the following idea: \emph{Can we train a neural network using some easily generated -- but rich in physics -- examples, and than apply the extracted knowledge to solve some more complex cases, not represented explicitly during the training?}
    
    Through the paper, our central focus will be on the following three aspects: \emph{knowledge extraction}, \emph{generalization capability}, and \emph{model interpretability}.
    To extract knowledge, we use the concept of \emph{curriculum learning} \cite{Bengio2009}. Namely, we construct a training set that allows a neural network to effectively learn the basic rules governing the physics of the quantum system.
    This procedure can also be viewed as \emph{knowledge distillation} \cite{Gou2021} from a physically-informed simulator, responsible for constructing the training examples, to an auxiliary network that learns from the prepared curriculum. This idea is rooted in the concept of teacher-network frameworks \cite{1412.6550}, whereby a smaller (and simpler) machine learning model is trained to approximate a larger, more complex system. However, whereas in the original formulation, this technique was  primarily used to reduce the final model complexity \cite{1503.02531}, i.e., in order to decrease the inference time and to reduce the overall computational and storage requirements, here we have another motivation. We want to promote the ability of the machine learning model to generalize (to make out-of-domain predictions). In other words, the goal here is to train on examples that are easy to construct, and then make predictions for cases that could potentially be difficult to simulate in a direct way.
    Finally, we want to observe how our machine learning-based emulator learns the essential physics from the physically-informed simulator. As we will show, by doing this, we can get additional insights about the nature of the underlying problem, discover symmetries, and measure the relative importance of the contributing features.
    
    In this work, as a prove of concept, we focus on the quantum dynamics of one-dimensional systems. While the problem is fairly easy to simulate in a traditional way \cite{Figueiras2018}, it also exhibits several non-trivial properties, such as  wave function inference, scattering, and tunnelling. Additionally, the emulator must learn to preserve the wave function normalization and must correctly interpret the real and imaginary part of the input.
    Another practical advantage of this problem formulation is that we can easily scale the difficulty of the task by analyzing potential landscapes of various complexity.

    The underlying motivation for focusing on quantum dynamics emulation tools is their use in simulating quantum systems and role in the design process of quantum devices, such as qubits and sensors \cite{2005.04681}. Specifically, modeling devices that are embedded in an environment requires challenging predictions of open quantum system dynamics \cite{DiCandia2015, Luchnikov2019}. Such simulations are inherently difficult on classical computers \cite{Loh1990, Prosen2007}. The reason is that \emph{direct} calculations can only be performed  for fairly small systems, as the limiting factor are the exponential dimensions of their Hilbert spaces \cite{Breuer2007}.
    Consequently, new tools that offer efficient and high-fidelity approximation of quantum dynamics can help the science community to model larger and more complex systems.

    In terms of related work, machine learning methods have recently been successfully used to solve many-electron Schr\"odinger equations \cite{Pfau2020, Hermann2020}. However, in contrast to our work, their focus was not on the quantum dynamics, but on finding equilibrium quantum states in electronic and molecular systems.
    Machine learning methods can be also used to solve partial differential equations (PDEs). In that context, some recent studies have been based on finite-dimensional approaches \cite{Zhu_2018, Bhatnagar_2019}, neural finite element \cite{smith2020eikonet, RAISSI2019686, raissi2018deep}, and Fourier neural operator methods \cite{lu2020deeponet, li2020neural}. However, in most of these approaches, the trained emulators can only generalize to a specific distribution of initial conditions. Consequently, they do not generalize in the space of the PDE parameters, and therefore they need to be re-trained for each new scenario.
    Machine learning was also used to emulate classical fluid dynamics \cite{sanchezgonzalez2020learning}. However, in those cases the focus was placed on accelerating  large-scale, classical simulations. In contrast, here we focus on quantum systems. Additionally, our training framework differs from a typical supervised setting that is often primarily concerned with in-domain predictions. We aim to extract knowledge from a curriculum of simple examples, and then generalize to more complex scenarios. Therefore our focus will be on training methods that facilitate generalization to (near) out-of-domain cases.
    
%%%%%%%%%%%%%%%%%%%%%%%%%%%%%%%%%%%%%%%%%%%%%%%%%%%%%%%%%%%%%%%%%:~

    \begin{figure*}
        \centering
        \includegraphics[width=1.0\textwidth]{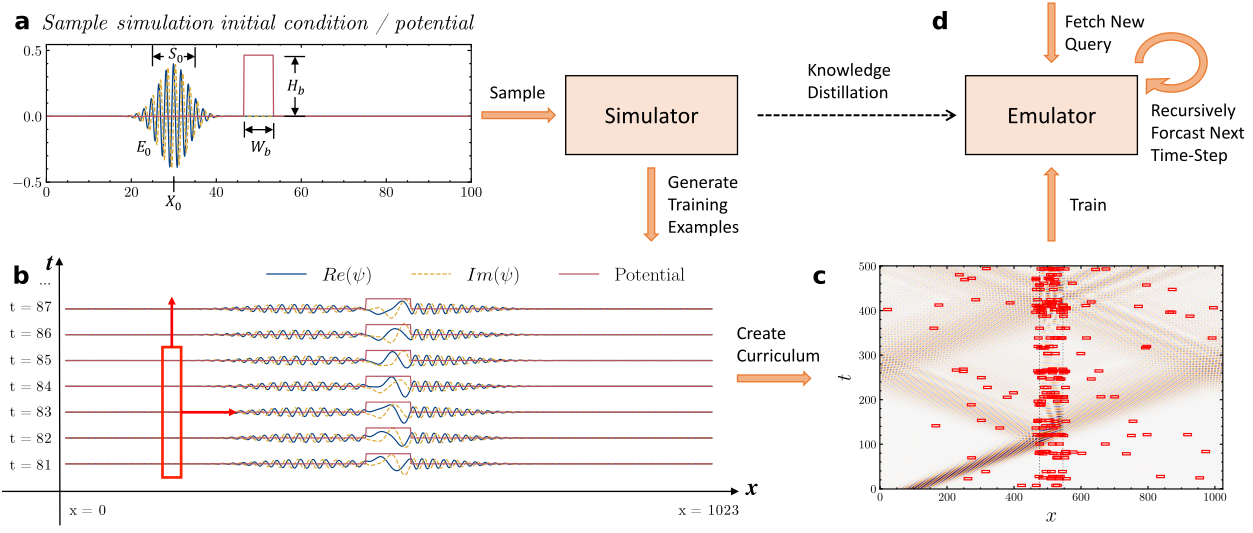}
        \caption{
            \textbf{Framework for Training a Machine Learning-Based Quantum Dynamics Emulator.}
            Illustration of how a machine learning-based emulator extracts knowledge from a physically informed simulator.
            {\bf a},~First, we sample initial conditions, according to which we generate a set of physical simulations. In this case, it is a set of time steps depicting Gaussian modulated quantum wave packets propagating through a system with a single rectangular potential barrier.
            {\bf b},~Next, by selecting small space and time slices (illustrated by the sliding red window), we construct individual training examples.
            {\bf c},~We sample from the space of all possible training examples to create a training curriculum. The idea is to select a diverse set of examples that illustrate all significant physical processes. In our case, these are quantum wave dispersion, tunnelling, scattering, and  interference.
            {\bf d}, Finally, the machine learning-based emulator learns from the prepared curriculum of training examples. We test the emulator by selecting novel cases and by recursively forecasting the next time steps of the system evolution.
        } 
        \label{fig:workflow}
    \end{figure*}

\section*{Results}

\subsection*{The Training Framework}

    In Figure~\ref{fig:workflow}, we show our proposed training framework for preparing machine learning-based quantum dynamics emulators.
    The main idea is based on the concepts of knowledge distillation \cite{Gou2021} and curriculum learning \cite{Bengio2009}. However, instead of extracting information from a larger machine learning model, our target is a simple, physically informed simulator.
    In detail, the framework consists of the following steps.
    First, the simulator samples the initial conditions and generates time-trajectories describing an evolution of the physical system of interest (cf.~Fig.~\ref{fig:workflow}{\bf a}).
    Next, we construct training examples from these recorded simulations (cf.~Fig.~\ref{fig:workflow}{\bf b}). 
    We select a diverse set of the training examples, making sure that all important phenomena of interest are represented (cf.~Fig.~\ref{fig:workflow}{\bf c}).
    This balanced curriculum of examples is consequently used as the input for the machine learning model. 
    We train the model, and then we validate it using some novel examples. When testing the model, we include cases that were not directly represented during the training, to measure whether the model can combine and generalize the acquired knowledge (cf.~Fig.~\ref{fig:workflow}{\bf d}).

\subsection*{Training and Testing Procedure}

    We illustrate our approach by training a neural network based emulator to predict the quantum dynamics of a one-dimensional system. To demonstrate that the emulator can extract knowledge from simple examples and generalize it in a non-trivial way, we \emph{restrict} the physical simulator to cases with only a single rectangular potential barrier. The emulator need to learn from these simple examples the basic properties of the wave function propagation, namely dispersion, scattering, tunneling, and quantum wave interference. Next, we test whether the emulator can  predict the time-evolution of wave packages in a more general case: e.g., for packages of different shapes propagating through an arbitrarily complex potential landscape.
    
    In detail, a single training simulation depicts a propagation of a Gaussian modulated wave packet. The initial conditions consists of: the center-of-mass position ($X_0$), the spread ($S_0$), and the energy ($E_0$) of the packet. 
    Due to the translational symmetry of the problem, we can  assume that the rectangular barrier is located at the center of the system, without any loss of generality. Consequently, the environment is fully described by two numbers: the height ($H_b$), and the width ($W_b$) of the rectangular barrier (cf.~Fig.~1{\bf a}, again; see also more details in the Supplementary Information).

    \begin{figure*}
        \begin{center}
        \includegraphics[width=\textwidth]{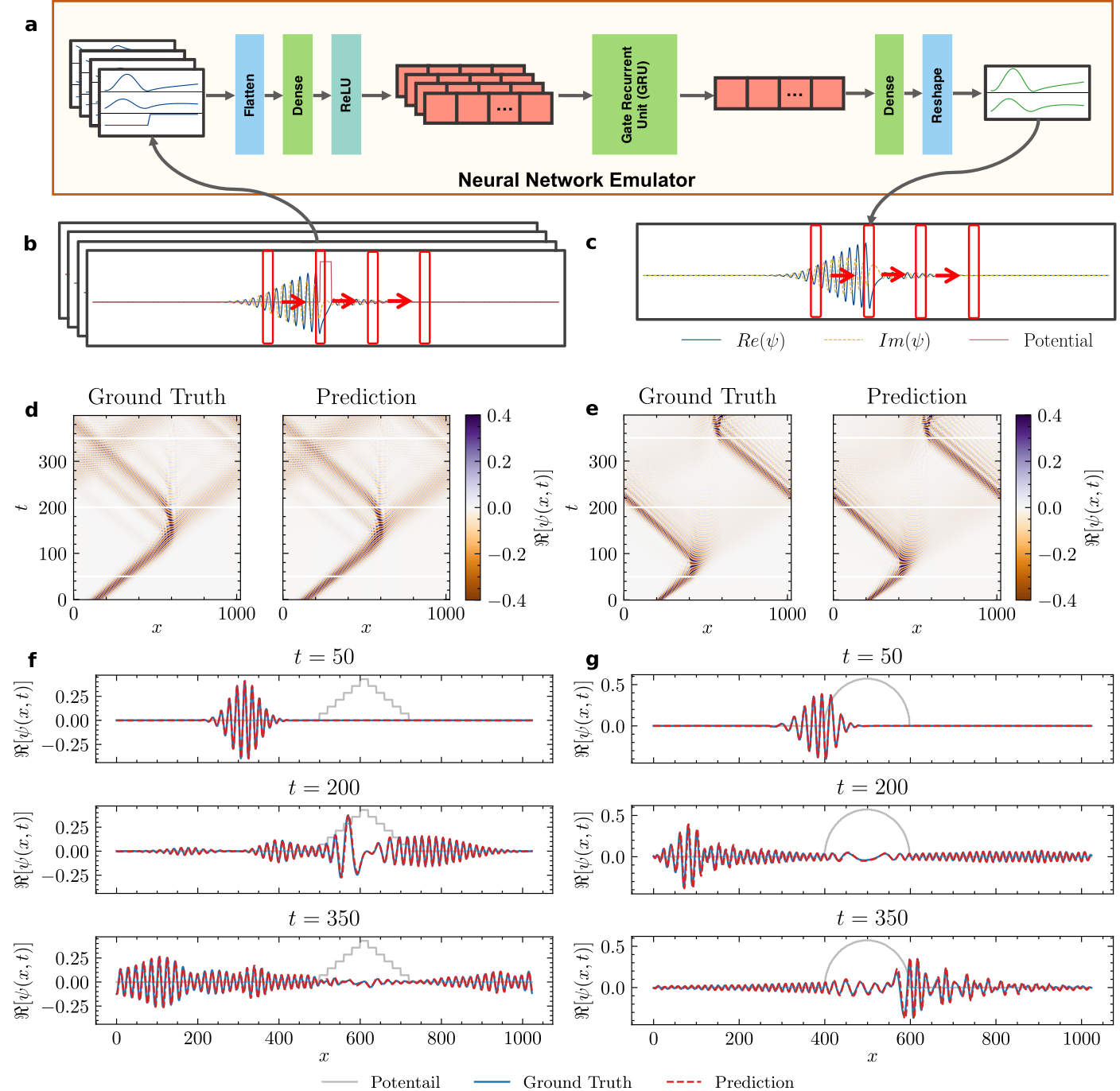}
        \end{center}
        \caption{\textbf{Machine Learning-Based Neural Network Emulator.} {\bf a}, Proposed neural network emulator architecture, along with the corresponding input and output. The rectangle boxes illustrate different types of neural network layers, and the red arrays represent the hidden states (hidden vectors). {\bf b}, Data representation of an exemplary raw input. The red rectangles represent spatial slices of the data (called \emph{windows} through the text). {\bf c}, Data representation of an exemplary output of the neural network emulator. {\bf d}, Ground truth and the predicted value of $\Re[\psi(x,t)]$ for an initial Gaussian wave packet and for a pyramid shaped potential barrier. {\bf f}, Comparisons of the predicted evolution and the ground truth for three temporal snapshots, $t \in \{50, 200, 350\}$, depicted by the white horizontal lines in panel {\bf d}.  {\bf e}, {\bf g}, A similar comparison, but for a half-circle shaped potential barrier instead. }  
        \label{fig:nn}
    \end{figure*}

\subsection*{The Architecture of the Quantum Dynamics Emulator}

    The architecture of our machine learning emulator is depicted in Figure~\ref{fig:nn}{\bf a}.
    We start by stacking four time steps of the original simulation data.
    To make the input independent of the size of the emulated system, we use windows of fixed width (represented by red rectangles in Fig.~\ref{fig:nn}{\bf b}).
    Each windowed chunk of data contains local information about the real and the imaginary part of the wave function as well as the local information about the potential landscape. Altogether, the data can be represented in the form of three channels (similar to the RGB channels when representing visual data), as depicted in on the left side of Fig.~\ref{fig:nn}{\bf a}.
    Next, we feed the input data to our neural network. The first few layers transform each time step into a hidden vector (represented graphically in our diagram by the red array).
    The role of the dense layer in our architecture is to allow the network to mix the information between different spatial points (non-local operation) and between channels (both local and non-local operations).
    In the next step, the hidden vectors pass through a set of gated recurrent units (GRU) \cite{cho2014learning}. These units are responsible for extracting  time-dependent information from the data.
    In the last step we reconstruct the wave function (the real and the imaginary part, represented by two channels, as depicted on the right side of Fig.~\ref{fig:nn}{\bf a}).

\subsection*{The Training Procedure}

    We generated a training data set with combinations of 189 different sets of initial Gaussian wave packets and 14 different rectangular potential barriers (in total, 2646 configurations). We kept the widths of the barrier fixed at $W_b = 7\, a.u.$ (71 pixels), the size of the environment at $N_x=1024$ pixels (with periodic boundary conditions enabled), and the window width at $W=23$ pixels.
    
    We trained the emulator for five epochs using the AdamW\cite{loshchilov2019decoupled} optimizer  with the MSE loss functions as the training objective.
    For more details regarding the architecture of our neural network-based emulator, the training procedure, and the composition of our training sets, see the Methods section.

\subsection*{Quantum Dynamics Emulation with Neural Networks}

    We present the results of the emulation and the comparison with the ground truth in Figs.~\ref{fig:nn}{\bf d}--{\bf g}.
    To show, that the emulator can generalize to out-of-domain situations, we have chosen two challenging shapes for the potential barrier: a step pyramid and a smooth half-circle. Since the emulator was trained only on a single rectangular barrier of a fixed width (much wider than the width of the step in the pyramid), those testing cases can not be reduced to any examples that were seen during the training.
    Instead, to make valid predictions, the neural networks must recombine the acquired knowledge in a non-trivial way.
    
    As we can see in Figs.~\ref{fig:nn}{\bf d}--{\bf g}, in both cases the predictions of our emulator match well the ground truth.
    It is worth of noticing, that the emulator makes its predictions in a recurrent manner (by using the predictions of the previous steps as input of the next step). Therefore, it is expected that predicting the evolution over hundreds of steps will cause some error accumulation. The fact that even after $350$ steps the error is negligible, speaks about the quality of the individual (step by step) predictions. This result indicates also, that long-term predictions of the dynamics are possible in our framework.

    As a conclusion, the proposed neural network emulator successfully learns the classical aspects of the wave dynamics, such as dispersion and interference. It also captures the more complex quantum phenomena, such as quantum tunneling.
    To further show that the emulator can generalize the acquired knowledge to make both in- and out-of-distribution predictions, we hand-designed a test data set with 12 freely dispersing cases, 11 rectangular barriers (with randomly chosen width and height), as well as 14 more challenging test cases depicting both multiple and irregularly shaped barriers (in total, 37 test instances). In all cases, the results were satisfactory, confirming the ability of the emulator to generalize to novel (and notably, more challenging) situations (see the detailed results in Figs.~\ref{fig:free_tests} and \ref{fig:barrier_tests} in the Supplementary Information).

\subsection*{Architecture Justification}
\label{sec:potential_results}

        \begin{figure*}
        \begin{center}
            \includegraphics[width=1.0\textwidth]{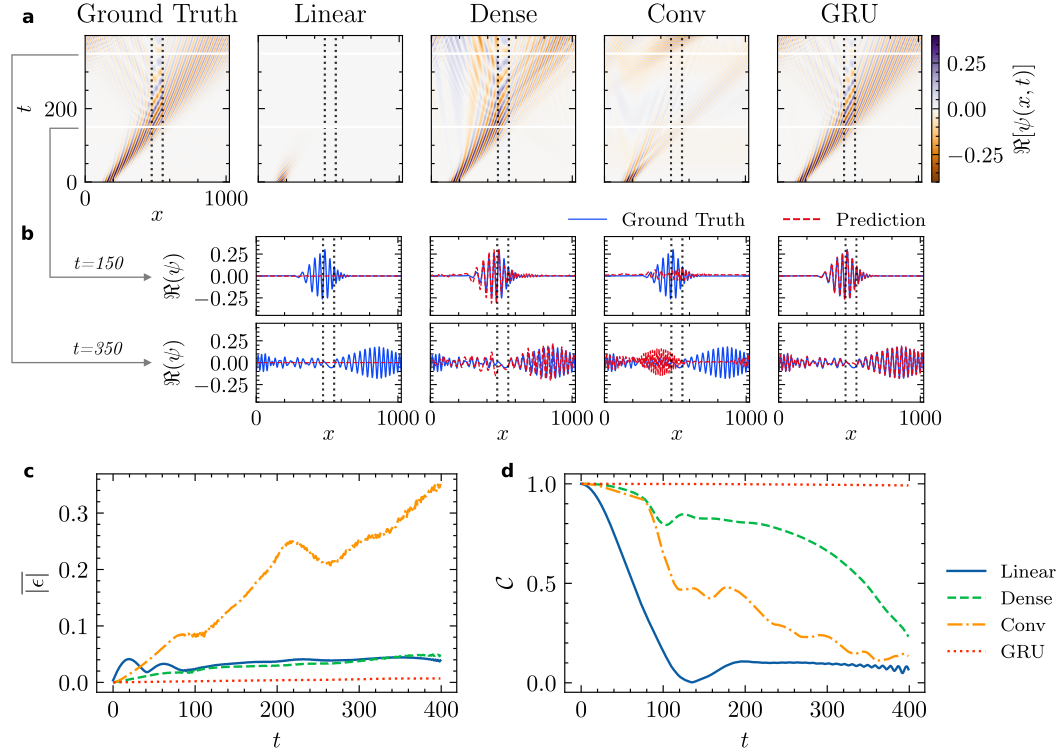}
        \end{center}
        \caption{\textbf{Scattering of a Quantum Wave Packet by a Rectangular Potential Barrier.} {\bf a}, Ground truth and the predicted value of $\Re[\psi(x,t)]$ for quantum wave packets scattered by a rectangular potential barrier. {\bf b}, Comparisons of emulated and exact solutions corresponding to two temporal snapshots (time steps 150 and 350 and depicted by the white horizontal lines in panel {\bf a}).
        %The solid blue line indicates the ground truth, while the dotted red line the predictions made using each emulator architecture.
        {\bf c} and {\bf d}, Time-step-by-time-step error propagation calculated as a mean absolute error, $\overline{|\epsilon|}$~(less is better), and normalized correlations, $\mathcal{C}$ (closer to $1$ is better), respectively.}
        \label{fig:barrier_one_example}
        \end{figure*} 
        
    In this section, we aim to provide some justification for the specific architecture of the emulator, that we introduced in the previous sections. 
    In order to verify, whether all the introduced complexity is necessary, we ask a question whether there exists a simpler machine learning model capable of learning quantum dynamics with a similar accuracy. Consequently, we compare our recurrent-network based approach with other popular network architectures.
    For consistency, in each case we use the same window-based scheme (to keep the input constant), while emulating the wave packet propagation for 400 time steps. When comparing with the ground truth, we measured the average mean absolute error (MAE) and the average normalized correlation~$\mathcal{C}$.

        \begin{table}
        \centering
        \setlength{\tabcolsep}{12pt}
        \caption{\textbf{Neural Network Architecture Comparison}.
        Performance comparison for different architectures of our machine learning-based emulator. As a metric, we used the mean absolute error $\overline{|\epsilon|}$ (less is better) and a normalized correlation $\mathcal{C}$ (closer to $1$ is better), both averaged over all spatial grid points, all time step, and all available test cases. 
        }
        \begin{tabularx}{\linewidth}{Xcll}
            \\\hline \hline
            Model  & Parameters & $\langle\overline{|\epsilon|}\rangle$  & $\langle\mathcal{C}\rangle$ \\ \hline \hline
            Linear & \phantom{0}3,220       &   0.0366 &   0.1597   \\ \hline
            Dense  & 27,163       &  0.0411  & 0.5729    \\ \hline
            Conv   &  28,889      &  0.1467 &  0.3667      \\ \hline
            GRU    &  40,204      &  {\bf 0.0051} &  {\bf 0.9953}        \\ \hline\hline
        \end{tabularx}
        \label{tab:barrier}
    \end{table}
    
    The results are presented in Table \ref{tab:barrier}, where the values represent the average performance over all our 37 test cases.
    For the comparison, we used three other models: a linear model, a densely connected feedforward model, and a convolutional model (for a detail description of each test case and each network architecture, see the Method section).
    As evident in our results, the proposed architecture (utilizing the gated recurrent units, GRU) outperforms all other, simpler architectures by a large margin.
    
    We present a details comparison for one of our test set in Figure~\ref{fig:barrier_one_example}.
    It is evident that the recurrent architecture provides the best results, whereas the simpler architectures fail to capture the complex long-time evolution of the wave packets.
    Notably, a failure of each simpler model can be used to justify different aspect of our final design. For example, the failure of the linear model might indicate the importance of the nonlinear activation functions included in our final model. The convolutional architecture captures quantum wave dispersion, but does not capture correctly the interaction between waves and the potential barrier. It might suggest, that to correctly capture the tunneling and scattering phenomena, we must mix the information not only between different spatial points but also between different channels.
    The dense architecture is able to capture both the reflection and the tunneling phenomena, but yields significant errors comparing to the recurrent-based model. This indicates, how important the temporal dimension is -- something what recurrent architectures are designed to explore, as they are capable of (selectively) storing the memory of the previous steps in their internal states -- and retrieving them when needed  \cite{cho2014learning, Gers1999}.

\subsection*{Generalization Capability}

    The usefulness of a neural network is mainly determined by its generalization capability.
    In this section, we provide a systematic analysis of both the in-domain and the out-of-domain predictive performance of our emulator.
    
    In the training process, the raw emulation data is broken up into small windows, and those windows are re-sampled to build the curriculum. One of the reason for doing so, was to balance cases featuring different distinct phenomena, e.g., free propagation vs.\ tunneling through the potential barrier. During the training, we artificially restricted our training instances only to those featuring Gaussian packets and single rectangular potential barriers. This allows us later to test the out-of-distribution generalizability potential -- namely, whether our emulator can correctly handle packets of different modulation or barriers of complex shape.
    
    The barrier used in our training instances had a~variable height, but a fixed width of $7\,a.u$. Notably, the width of the window, that we used to cut chunks of the input data for our neural network, was $2.25\,a.u.$ -- i.e., it was smaller than the width of the potential barrier itself. As a result, the network was exposed during the training to three distinct situations: (1)~freely dispersing quantum waves in zero potential; (2)~quantum wave propagation with non-zero constant potential; (3)~quantum wave propagation with a potential step from zero to a constant value of the potential (or vice versa).
    While it is obvious that our emulator should handle rectangular potential barriers wider than the width of the window, since such situations are analogous to those already encountered in the training process, it is not all so obvious that the same should happen with barriers of a smaller width, yet alone with barriers of different shapes than rectangular.
    However, as it was already presented in Fig.~\ref{fig:nn}, those cases do not present a challenge for our emulator.

        \begin{figure*}
        \begin{center}
            \includegraphics[width=\textwidth]{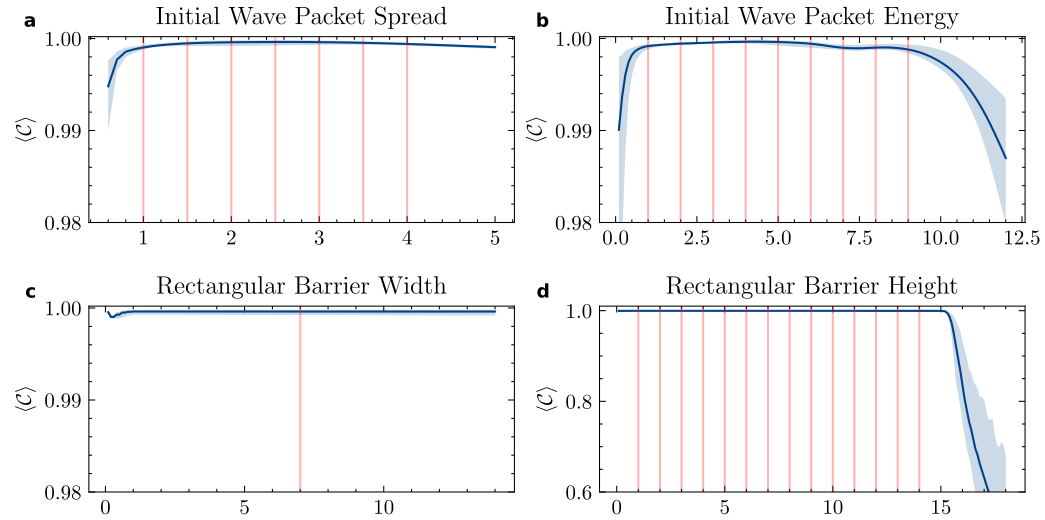}
        \end{center}
         \caption{\textbf{Generalization of Simulation Parameters.}  Average normalized correlations, denoted by $\langle \mathcal{C} \rangle$, as functions of four simulation parameters. {\bf a}, The initial wave packet spread. {\bf b}, The initial wave packet energy. {\bf c}, The rectangular barrier width. {\bf d}, The rectangular barrier height. We repeated each experiments using five different random seeds. The bold blue line in each panel shows the average normalized correlation, and the blue shades indicates the span between the minimum and maximum values recorded in those five trials. The red vertical lines denote the training values for each emulation parameter.} 
        \label{fig:generalization}
        \end{figure*}

    In Figure~\ref{fig:generalization}, we present a more systematic study of that phenomena. We have evaluated how the accuracy of the emulation is affected when changing the following parameters: (a) the initial wave packet spread, (b) the initial wave packet energy, (c) the rectangular barrier width, and (d) the rectangular barrier height. In the experiments, we alter one of the parameters, while holding all other parameters at a constant value. We report the average correlation calculated across all the spatial grid points and all the emulated time steps.
    As we can see, the performance of the emulator remains effectively unchanged for all the tested values of the initial wave spread and the rectangular barrier width. This demonstrates that the network generalizes well with respect to those parameters. Remarkably, the performance does not deteriorate even when the barrier width becomes smaller than the size of the window, i.e., $2.5\,a.u$. This explains why our emulator was able to correctly emulate the evolution with presence of continuously-shaped barriers (that, due to the discrete nature of our input, can be seen as a collection of adjacent 1-pixel-wide rectangular barriers).

    With respect to the value of the initial wave packet energy, the correlation curve has a ``$\cap$'' shape, indicating that the accuracy deteriorates as we leave the energy-region sampled during the training stage. However, this is as much a failure to generalize as the limitation of the way how we represent the data and the environment. Lower energy of the packet means that the evolution of the quantum wave is slower -- and we have to predict more steps of the simulation to cover the same spatial distance as for wave packets with a higher energy. Due to the recurrent process of our emulation, this means larger error accumulation. On the other end, when the energy is high, the wave functions change rapidly within each discretized time unit. This means a larger potential for error each time we predict the next time step.
    Similar situation also happens when the barrier height increases. A steeper barrier means
    a more dramatic changes to the values of the wave function happening between two consequent time steps. Those both effects can be mitigated by increasing either the spatial or temporal resolution of our emulation.

\subsection*{Model Interpretability Through Input Feature Attribution}

        \begin{figure*}
        \begin{center}
            \includegraphics[width=\textwidth]{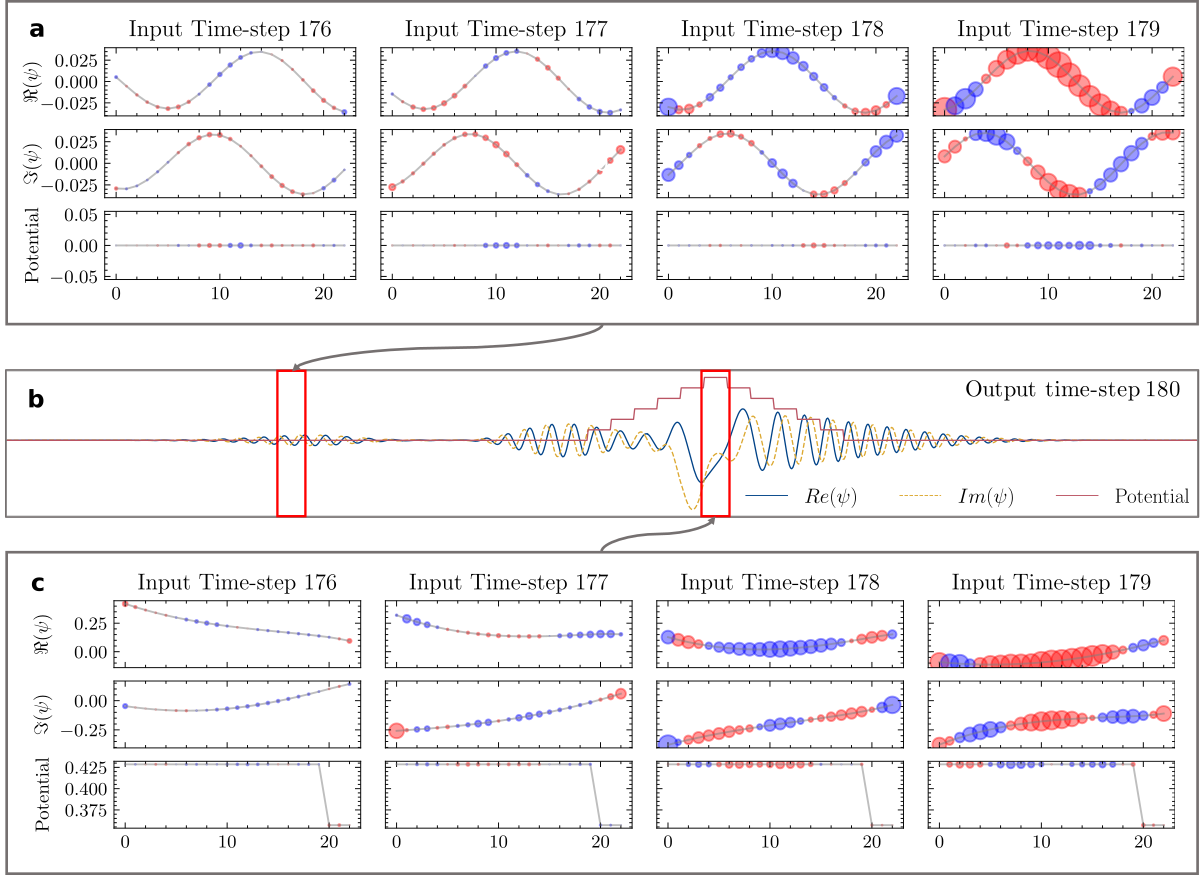}
        \end{center}
         \caption[Direct-gradient examples.]{{\bf Model Interpretability.} {\bf a}, Direct gradients calculated for a freely dispersing quantum wave (far from any potential barrier). To present the results, we overlaid the values of the gradient with the input pixels, with red circles denoting positive direct-gradient values, and with blue circles denoting negative values.
         {\bf b}, Two wave functions: one far from any potential barrier (left); second interacting with a step pyramid-shaped potential barrier (right).
         {\bf c}, Direct gradients calculated for a quantum wave interacting with a potential barrier.}
        \label{fig:dg}
        \end{figure*}

    In this section, to get some insight into how our emulator makes the prediction, we measure the feature attribution (the contribution of each data input to the final prediction).
    We employ the direct gradient method. Namely, for a given target pixel in the output window, we calculate its gradients with respect to every input value.
    We follow the intuition that larger gradient is associated with higher importance of measured input \cite{Simonyan2013}.
    Since all the target values are complex numbers, this method gives us two sets of values, one for the real and one for the imaginary value of the target pixel,
    \begin{equation}
        \frac{\partial \Re\left(\psi_{\text {target}}\right)}{\partial\{\ldots\}}
        \hspace{12pt} \mbox{and} \hspace{12pt}
        \frac{\partial \Im\left(\psi_{\text {target}}\right)}{\partial\{\ldots\}}.
    \end{equation}
    
    In Figure~\ref{fig:dg}, we show a sample of the results.
    Namely, we selected one test example, and as the target we took the central pixel of the predicted frame. Next, we calculated the gradient with respect to each input value.
    In this concrete example, we can see that the importance of the input pixels increases with each consecutive time steps. However, how much exactly the early steps matter depends on the region where the predictions are made.
    Far from the potential barrier (cf.~Fig.~\ref{fig:dg}{\bf a}), the gradient measured with respect to the pixels of the first two steps (no.~176 and 177) is close to zero, indicating that those steps carry relative little importance.
    However, when predicting the evolution of the wave function in the vicinity of the potential barrier (cf.~Fig.~\ref{fig:dg}{\bf c}), the importance of the third-from-the last step (no.~177) increases noticeable.
    This indicates, that when predicting free propagation, the network simply performs a linear extrapolation based on the latest two time-steps. However, when predicting tunneling and scattering, the emulator reaches beyond the last two time-steps for additional information, likely due to the non-linear character of those predictions.
    
    %For more detail regarding the specifics of the network architecture, see the Supplementary Information. Among others, to verify the above interpretation, we tested there the performance of the emulator while changing the number of available time-steps from $1$ to $8$.
    %The results confirm, that as evident in the analysed example, access to a longer history is important only as long as interaction with a potential barrier is considered.

%%%%%%%%%%%%%%%%%%%%%%%%%%%%%%%%%%%%%%%%%%%%%%%%%%%%%%%%%%%%%%%%%:~

\section*{Discussion}

    Interpreting neural networks creates interesting possibilities.
    Namely, by analysing how machine learning models make their predictions, we can learn new facts about the underlying physical problem.
    As an example, using the direct gradient descend method, despite its simplicity and known limitations\cite{Sundararajan2017}, we can discover interesting relations between the input parameters.
    By analyzing several examples like those depicted in Fig.~\ref{fig:dg}, we can observe a strong relationship between the averaged direct gradients of $\Re(\psi_{\text{target}})$ and $\Im(\psi_{\text{target}})$, namely,
    \begin{eqnarray}
        \frac{\partial \Re(\psi_{\text{target}})}{\partial \Re(\psi_{\text{input}})} &=&  \frac{\partial \Im(\psi_{\text{target}})}{\partial \Im(\psi_{\text{input}})},
        \hspace{12pt}\mbox{and}\\[6pt]
        \frac{\partial \Re(\psi_{\text{target}})}{\partial \Im(\psi_{\text{input}})} &=& -\frac{\partial \Im(\psi_{\text{target}})}{\partial \Re(\psi_{\text{input}})}.
    \end{eqnarray}
    This result can be used to (re)discover the existence of Cauchy-Riemann relations\cite{Riemann1851}.
    Notable, this happens, despite the fact that we have not imposed any prior knowledge of wave dynamics or complex analysis during the training procedure.
    Our model was able to learn the correct relation directly from the presented training examples.
    
    This trend, of looking at machine learning algorithms as something more than just black-box systems capable of extracting patterns from the data or solving given classification and regression problems, opens new possibilities \cite{Zdeborov2017, Zdeborov2020}.
    As it is evident in our example, we can use machine learning to get a better insight into the studied problem. 
    There is an increasing number of papers, exploring this direction.
    For example, V.~Bapst et al., in their article\cite{Bapst2020}, demonstrated how to train a graph neural network to predict the time-evolution of a glass system. By measuring the contributions to the prediction from particles located on subsequent ``shells'' around the target particle, they were able to estimate the effective correlation length of the interactions.
    They were also able to qualitatively show, that when system approaches the glass transition, the effective cut-off distance for the particle-particle interactions increases rapidly, a phenomena that is experimentally observed in glass systems.
    As an another example, Lee et al., in their work\cite{Lee2022}, trained support vector machines to predict the preferred phase type for the multi-principal element alloys. Next, by interpreting the trained models, they were able to discover phase-specific relations between the chemical composition of the alloys and their experimentally observed phases. Knowing what influence the phase formation, can be important both in the scientific context, as it increases our understanding, and can help us to refine our theoretical models, as well as from the manufacturing perspective, as it enables synthesis of materials with desired mechanical properties.
    
    %of using machine learning for scientific concept discovery. % machine learning-enabled scientific concept discovery
    
    Another topic discussed in our paper is the ability of machine learning models to generalize to out-of-distribution examples. For the completeness, it is important to discuss what we mean by this term. We train our model on some simple examples, generated by an emulator restricted to a specific region of the parameter space.
    Next, we wanted to generalize \emph{in that parameter space}. Namely, our goal is to predict the time-evolution of the system for parameters that are outside of the scope used during the training.
    This is in contrast to the typical understanding of generalization, as the ability of handling a novel (unseen during the training) data-input instances, that nevertheless come from the same distribution as the training examples.
    While our parameter space is relatively low-dimensional (we have just a couple of variables describing the initial conditions of the system), it is not the case for the ambient space of the inputs (each input instance is a tensor of $4 \times 23 \times 3$, this is $276$ dimensions).
    This high-dimensionality of the ambient space has profound implications.
    Balestriero,  Pesenti and LeCun, in their work\cite{Balestriero2021}, have shown that in highly dimensional space all predictions are effectively extrapolations, not interpolations. The logic is, that the number of training examples required for a prediction to be seen as interpolation, grows at least as fast as exponentially with the increasing dimension of the input. Effectively, for any dimensions above a few dozens, the training points become very sparse. Therefore, any point during the testing phase is likely to be far (at least along some dimensions) from any other point observed during the training. This makes that almost every testing point is located outside of the convex hull defined by the training examples -- thus predictions of those points is nothing else than an extrapolation.
    In our case, however, we are interested in the ability to extrapolate in the space of the original parameter space, that control our physic-informed emulator, not in the ambient space of the neural network input. Since our parameters space has much lower dimension -- just below 10 -- thus, we can distinguish cases that require interpolation and extrapolation.
    As an example, if our model was trained by showing propagation of wave packages with the energy spread $S_0 \in \{1, 1.5, 2, 2.5, 3, 3.5, 4\}$, then to predict an evolution for a package that has spread $2.7$ would be an interpolation, and to predict the evolution for the spread $5$ would be an extrapolation.
    In Fig.~\ref{fig:generalization}, we have shown the performance of our emulator both in the interpolation and the extrapolation regime. We have also discussed, that the emulator can generalize outside the explored region of the parameter space, at lest along some dimensions. 
    
    There are other works discussing generalization to out-of-domain cases\cite{Wald2021, Wang2021}. However, in those works the question of generalizability was asked mostly in the context of out-of-distribution detection, and used mostly in the context of fraud detection\cite{Pang2021}, general novelty detection\cite{Yang2021} or as a means for increasing trustworthiness of machine learning systems\cite{Liu2020}. In this paper, we explored this concepts in the context of obtaining robust emulators of physical systems.

\section*{Conclusions}

    We have designed a training framework that allows us to extract knowledge from a restricted physically informed simulator.
    Furthermore, we  have demonstrated how the interpretation of machine learning models can lead to scientific concept discovery.
    We also showed that it is possible for the network to train on simple examples and generalize to  more general cases. 
    
    In future work we will optimize training schemes that maximize the generalization capability. Moreover, we will apply this methodology to physically challenging systems, e.g., including interactions. Furthermore, we will implement more sophisticated interpretability techniques and increase the robustness of the emulator by utilizing  known symmetries, e.g., by constructing physically informed neural networks. Finally, we would like to pursue an iterative framework, whereby any time a new symmetry is discovered, it is implemented  as a system constraint, repeating this procedure for a full characterization of the underlying physical system.

    %We have designed a training framework that allows to extract knowledge from a restricted physically-informed simulator. We also have show, that interpretation of machine learning models can lead to scientific concept discovery.
    
    %The nature of the work here was to show the potential direction of research. Namely, our goal was to show that it is possible to train on some simple examples and generalize the knowledge to predict more general cases. We has deliberately chosen a simple system, to deliver a clear proof of concept. We hope to encourage people to follow this direction. Future work might include: search for training schemes that allows to increase the generalization capability. Extending our discussion to more challenging systems. Implementation of more sophisticated interpretability techniques. Increasing the robustness of the emulator by utilizing the known symmetries (e.g., by constructing physically informed neural networks). It might also be a promising direction to propose an iterative framework, where any time we discover new symmetries, we implement them as constraint in the system, and than we repeat the process of searching for new symmetries again.

%%%%%%%%%%%%%%%%%%%%%%%%%%%%%%%%%%%%%%%%%%%%%%%%%%%%%%%%%%%%%%%%%:~

%\section*{References}% Un-comment me after we copy here the output of the .bbl file. 
\bibliographystyle{naturemag}  % Nature Bibtex Style
\def\bibsection{\section*{\refname}} 
\bibliography{bibliography.bib}

\begin{thebibliography}{10}
\expandafter\ifx\csname url\endcsname\relax
  \def\url#1{\texttt{#1}}\fi
\expandafter\ifx\csname urlprefix\endcsname\relax\def\urlprefix{URL }\fi
\providecommand{\bibinfo}[2]{#2}
\providecommand{\eprint}[2][]{\url{#2}}

\bibitem{Bahri2020}
\bibinfo{author}{Bahri, Y.} \emph{et~al.}
\newblock \bibinfo{title}{Statistical mechanics of deep learning}.
\newblock \emph{\bibinfo{journal}{Annual Review of Condensed Matter Physics}}
  \textbf{\bibinfo{volume}{11}}, \bibinfo{pages}{501--528}
  (\bibinfo{year}{2020}).
\newblock
  \urlprefix\url{https://doi.org/10.1146/annurev-conmatphys-031119-050745}.

\bibitem{Hornik1989}
\bibinfo{author}{Hornik, K.}, \bibinfo{author}{Stinchcombe, M.} \&
  \bibinfo{author}{White, H.}
\newblock \bibinfo{title}{Multilayer feedforward networks are universal
  approximators}.
\newblock \emph{\bibinfo{journal}{Neural Networks}}
  \textbf{\bibinfo{volume}{2}}, \bibinfo{pages}{359--366}
  (\bibinfo{year}{1989}).
\newblock \urlprefix\url{https://doi.org/10.1016/0893-6080(89)90020-8}.

\bibitem{Delalleau2011}
\bibinfo{author}{Delalleau, O.} \& \bibinfo{author}{Bengio, Y.}
\newblock \bibinfo{title}{Shallow vs. deep sum-product networks}.
\newblock In \bibinfo{editor}{Shawe-Taylor, J.}, \bibinfo{editor}{Zemel, R.},
  \bibinfo{editor}{Bartlett, P.}, \bibinfo{editor}{Pereira, F.} \&
  \bibinfo{editor}{Weinberger, K.~Q.} (eds.) \emph{\bibinfo{booktitle}{Advances
  in Neural Information Processing Systems}}, vol.~\bibinfo{volume}{24}
  (\bibinfo{publisher}{Curran Associates, Inc.}, \bibinfo{year}{2011}).
\newblock
  \urlprefix\url{https://proceedings.neurips.cc/paper/2011/file/8e6b42f1644ecb1327dc03ab345e618b-Paper.pdf}.

\bibitem{Poole2016}
\bibinfo{author}{Poole, B.}, \bibinfo{author}{Lahiri, S.},
  \bibinfo{author}{Raghu, M.}, \bibinfo{author}{Sohl-Dickstein, J.} \&
  \bibinfo{author}{Ganguli, S.}
\newblock \bibinfo{title}{Exponential expressivity in deep neural networks
  through transient chaos}.
\newblock In \bibinfo{editor}{Lee, D.}, \bibinfo{editor}{Sugiyama, M.},
  \bibinfo{editor}{Luxburg, U.}, \bibinfo{editor}{Guyon, I.} \&
  \bibinfo{editor}{Garnett, R.} (eds.) \emph{\bibinfo{booktitle}{Advances in
  Neural Information Processing Systems}}, vol.~\bibinfo{volume}{29}
  (\bibinfo{publisher}{Curran Associates, Inc.}, \bibinfo{year}{2016}).
\newblock
  \urlprefix\url{https://proceedings.neurips.cc/paper/2016/file/148510031349642de5ca0c544f31b2ef-Paper.pdf}.

\bibitem{LeCun2015}
\bibinfo{author}{LeCun, Y.}, \bibinfo{author}{Bengio, Y.} \&
  \bibinfo{author}{Hinton, G.}
\newblock \bibinfo{title}{Deep learning}.
\newblock \emph{\bibinfo{journal}{Nature}} \textbf{\bibinfo{volume}{521}},
  \bibinfo{pages}{436--444} (\bibinfo{year}{2015}).
\newblock \urlprefix\url{https://doi.org/10.1038/nature14539}.

\bibitem{Bau2020}
\bibinfo{author}{Bau, D.} \emph{et~al.}
\newblock \bibinfo{title}{Understanding the role of individual units in a deep
  neural network}.
\newblock \emph{\bibinfo{journal}{Proceedings of the National Academy of
  Sciences}} \textbf{\bibinfo{volume}{117}}, \bibinfo{pages}{30071--30078}
  (\bibinfo{year}{2020}).
\newblock \urlprefix\url{https://doi.org/10.1073/pnas.1907375117}.

\bibitem{Sengupta2020}
\bibinfo{author}{Sengupta, S.} \emph{et~al.}
\newblock \bibinfo{title}{A review of deep learning with special emphasis on
  architectures, applications and recent trends}.
\newblock \emph{\bibinfo{journal}{Knowledge-Based Systems}}
  \textbf{\bibinfo{volume}{194}}, \bibinfo{pages}{105596}
  (\bibinfo{year}{2020}).
\newblock \urlprefix\url{https://doi.org/10.1016/j.knosys.2020.105596}.

\bibitem{Jiang2017}
\bibinfo{author}{Jiang, F.} \emph{et~al.}
\newblock \bibinfo{title}{Artificial intelligence in healthcare: past, present
  and future}.
\newblock \emph{\bibinfo{journal}{Stroke and Vascular Neurology}}
  \textbf{\bibinfo{volume}{2}}, \bibinfo{pages}{230--243}
  (\bibinfo{year}{2017}).
\newblock \urlprefix\url{https://doi.org/10.1136/svn-2017-000101}.

\bibitem{great2018ai}
\bibinfo{author}{{House of Lords: Select Committee on Artificial
  Intelligence}}.
\newblock \emph{\bibinfo{title}{AI in the UK: Ready, Willing and Able? Report
  of Session 2017-19.}}
\newblock No. \bibinfo{number}{100} in \bibinfo{series}{HL paper}
  (\bibinfo{publisher}{Dandy Booksellers Limited}, \bibinfo{year}{2018}).
\newblock \urlprefix\url{https://books.google.com/books?id=sNyMtgEACAAJ}.

\bibitem{carleo2019machine}
\bibinfo{author}{Carleo, G.} \emph{et~al.}
\newblock \bibinfo{title}{Machine learning and the physical sciences}.
\newblock \emph{\bibinfo{journal}{Rev. Mod. Phys.}}
  \textbf{\bibinfo{volume}{91}}, \bibinfo{pages}{045002}
  (\bibinfo{year}{2019}).
\newblock
  \urlprefix\url{https://link.aps.org/doi/10.1103/RevModPhys.91.045002}.

\bibitem{Schmidt2019}
\bibinfo{author}{Schmidt, J.}, \bibinfo{author}{Marques, M. R.~G.},
  \bibinfo{author}{Botti, S.} \& \bibinfo{author}{Marques, M. A.~L.}
\newblock \bibinfo{title}{Recent advances and applications of machine learning
  in solid-state materials science}.
\newblock \emph{\bibinfo{journal}{npj Computational Materials}}
  \textbf{\bibinfo{volume}{5}} (\bibinfo{year}{2019}).
\newblock \urlprefix\url{https://doi.org/10.1038/s41524-019-0221-0}.

\bibitem{Carrasquilla2020}
\bibinfo{author}{Carrasquilla, J.}
\newblock \bibinfo{title}{Machine learning for quantum matter}.
\newblock \emph{\bibinfo{journal}{Advances in Physics: X}}
  \textbf{\bibinfo{volume}{5}}, \bibinfo{pages}{1797528}
  (\bibinfo{year}{2020}).
\newblock \urlprefix\url{https://doi.org/10.1080/23746149.2020.1797528}.

\bibitem{Alzubaidi2021}
\bibinfo{author}{Alzubaidi, L.} \emph{et~al.}
\newblock \bibinfo{title}{Review of deep learning: concepts, {CNN}
  architectures, challenges, applications, future directions}.
\newblock \emph{\bibinfo{journal}{Journal of Big Data}}
  \textbf{\bibinfo{volume}{8}} (\bibinfo{year}{2021}).
\newblock \urlprefix\url{https://doi.org/10.1186/s40537-021-00444-8}.

\bibitem{2008.03703}
\bibinfo{author}{Feldman, V.} \& \bibinfo{author}{Zhang, C.}
\newblock \bibinfo{title}{What neural networks memorize and why: Discovering
  the long tail via influence estimation}.
\newblock \emph{\bibinfo{journal}{arXiv}} \textbf{\bibinfo{volume}{2008.03703}}
  (\bibinfo{year}{2020}).
\newblock \urlprefix\url{https://arxiv.org/abs/2008.03703}.

\bibitem{Chai2020}
\bibinfo{author}{Chai, X.} \emph{et~al.}
\newblock \bibinfo{title}{Deep learning for irregularly and regularly missing
  data reconstruction}.
\newblock \emph{\bibinfo{journal}{Scientific Reports}}
  \textbf{\bibinfo{volume}{10}} (\bibinfo{year}{2020}).
\newblock \urlprefix\url{https://doi.org/10.1038/s41598-020-59801-x}.

\bibitem{Haley1992}
\bibinfo{author}{Haley, P.} \& \bibinfo{author}{Soloway, D.}
\newblock \bibinfo{title}{Extrapolation limitations of multilayer feedforward
  neural networks}.
\newblock In \emph{\bibinfo{booktitle}{Proceedings of the {IJCNN} International
  Joint Conference on Neural Networks}} (\bibinfo{publisher}{{IEEE}},
  \bibinfo{year}{1992}).
\newblock \urlprefix\url{https://doi.org/10.1109/ijcnn.1992.227294}.

\bibitem{2009.11848}
\bibinfo{author}{Xu, K.} \emph{et~al.}
\newblock \bibinfo{title}{How neural networks extrapolate: From feedforward to
  graph neural networks}.
\newblock In \emph{\bibinfo{booktitle}{9th International Conference on Learning
  Representations, {ICLR} 2021, Virtual Event, Austria, May 3-7, 2021}}
  (\bibinfo{publisher}{OpenReview.net}, \bibinfo{year}{2021}).
\newblock \urlprefix\url{https://openreview.net/forum?id=UH-cmocLJC}.

\bibitem{Han2018}
\bibinfo{author}{Han, J.}, \bibinfo{author}{Jentzen, A.} \& \bibinfo{author}{E,
  W.}
\newblock \bibinfo{title}{Solving high-dimensional partial differential
  equations using deep learning}.
\newblock \emph{\bibinfo{journal}{Proceedings of the National Academy of
  Sciences}} \textbf{\bibinfo{volume}{115}}, \bibinfo{pages}{8505--8510}
  (\bibinfo{year}{2018}).
\newblock \urlprefix\url{https://doi.org/10.1073/pnas.1718942115}.

\bibitem{Karniadakis2021}
\bibinfo{author}{Karniadakis, G.~E.} \emph{et~al.}
\newblock \bibinfo{title}{Physics-informed machine learning}.
\newblock \emph{\bibinfo{journal}{Nature Reviews Physics}}
  \textbf{\bibinfo{volume}{3}}, \bibinfo{pages}{422--440}
  (\bibinfo{year}{2021}).
\newblock \urlprefix\url{https://doi.org/10.1038/s42254-021-00314-5}.

\bibitem{Iten2020}
\bibinfo{author}{Iten, R.}, \bibinfo{author}{Metger, T.},
  \bibinfo{author}{Wilming, H.}, \bibinfo{author}{del Rio, L.} \&
  \bibinfo{author}{Renner, R.}
\newblock \bibinfo{title}{Discovering physical concepts with neural networks}.
\newblock \emph{\bibinfo{journal}{Physical Review Letters}}
  \textbf{\bibinfo{volume}{124}} (\bibinfo{year}{2020}).
\newblock \urlprefix\url{https://doi.org/10.1103/physrevlett.124.010508}.

\bibitem{Murdoch2019}
\bibinfo{author}{Murdoch, W.~J.}, \bibinfo{author}{Singh, C.},
  \bibinfo{author}{Kumbier, K.}, \bibinfo{author}{Abbasi-Asl, R.} \&
  \bibinfo{author}{Yu, B.}
\newblock \bibinfo{title}{Definitions, methods, and applications in
  interpretable machine learning}.
\newblock \emph{\bibinfo{journal}{Proceedings of the National Academy of
  Sciences}} \textbf{\bibinfo{volume}{116}}, \bibinfo{pages}{22071--22080}
  (\bibinfo{year}{2019}).
\newblock \urlprefix\url{https://doi.org/10.1073/pnas.1900654116}.

\bibitem{Joshi2021}
\bibinfo{author}{Joshi, G.}, \bibinfo{author}{Walambe, R.} \&
  \bibinfo{author}{Kotecha, K.}
\newblock \bibinfo{title}{A review on explainability in multimodal deep neural
  nets}.
\newblock \emph{\bibinfo{journal}{{IEEE} Access}} \textbf{\bibinfo{volume}{9}},
  \bibinfo{pages}{59800--59821} (\bibinfo{year}{2021}).
\newblock \urlprefix\url{https://doi.org/10.1109/access.2021.3070212}.

\bibitem{Bengio2009}
\bibinfo{author}{Bengio, Y.}, \bibinfo{author}{Louradour, J.},
  \bibinfo{author}{Collobert, R.} \& \bibinfo{author}{Weston, J.}
\newblock \bibinfo{title}{Curriculum learning}.
\newblock In \emph{\bibinfo{booktitle}{Proceedings of the 26th Annual
  International Conference on Machine Learning - {ICML} {\textquotesingle}09}}
  (\bibinfo{publisher}{{ACM} Press}, \bibinfo{year}{2009}).
\newblock \urlprefix\url{https://doi.org/10.1145/1553374.1553380}.

\bibitem{Gou2021}
\bibinfo{author}{Gou, J.}, \bibinfo{author}{Yu, B.}, \bibinfo{author}{Maybank,
  S.~J.} \& \bibinfo{author}{Tao, D.}
\newblock \bibinfo{title}{Knowledge distillation: A survey}.
\newblock \emph{\bibinfo{journal}{International Journal of Computer Vision}}
  \textbf{\bibinfo{volume}{129}}, \bibinfo{pages}{1789--1819}
  (\bibinfo{year}{2021}).
\newblock \urlprefix\url{https://doi.org/10.1007/s11263-021-01453-z}.

\bibitem{1412.6550}
\bibinfo{author}{Romero, A.} \emph{et~al.}
\newblock \bibinfo{title}{Fitnets: Hints for thin deep nets}.
\newblock In \bibinfo{editor}{Bengio, Y.} \& \bibinfo{editor}{LeCun, Y.} (eds.)
  \emph{\bibinfo{booktitle}{3rd International Conference on Learning
  Representations, {ICLR} 2015, San Diego, CA, USA, May 7-9, 2015, Conference
  Track Proceedings}} (\bibinfo{year}{2015}).
\newblock \urlprefix\url{http://arxiv.org/abs/1412.6550}.

\bibitem{1503.02531}
\bibinfo{author}{Hinton, G.}, \bibinfo{author}{Vinyals, O.} \&
  \bibinfo{author}{Dean, J.}
\newblock \bibinfo{title}{Distilling the knowledge in a neural network}.
\newblock In \emph{\bibinfo{booktitle}{NIPS Deep Learning and Representation
  Learning Workshop}} (\bibinfo{year}{2015}).
\newblock \urlprefix\url{http://arxiv.org/abs/1503.02531}.

\bibitem{Figueiras2018}
\bibinfo{author}{Figueiras, E.}, \bibinfo{author}{Olivieri, D.},
  \bibinfo{author}{Paredes, A.} \& \bibinfo{author}{Michinel, H.}
\newblock \bibinfo{title}{An open source virtual laboratory for the
  schr\"{o}dinger equation}.
\newblock \emph{\bibinfo{journal}{European Journal of Physics}}
  \textbf{\bibinfo{volume}{39}}, \bibinfo{pages}{055802}
  (\bibinfo{year}{2018}).
\newblock \urlprefix\url{https://doi.org/10.1088/1361-6404/aac999}.

\bibitem{2005.04681}
\bibinfo{author}{Meyerov, I.}, \bibinfo{author}{Liniov, A.},
  \bibinfo{author}{Ivanchenko, M.} \& \bibinfo{author}{Denisov, S.}
\newblock \bibinfo{title}{Simulating quantum dynamics: Evolution of algorithms
  in the hpc context}.
\newblock \emph{\bibinfo{journal}{arXiv}} \textbf{\bibinfo{volume}{2005.04681}}
  (\bibinfo{year}{2020}).
\newblock \urlprefix\url{https://arxiv.org/abs/2005.04681}.

\bibitem{DiCandia2015}
\bibinfo{author}{Candia, R.~D.}, \bibinfo{author}{Pedernales, J.~S.},
  \bibinfo{author}{del Campo, A.}, \bibinfo{author}{Solano, E.} \&
  \bibinfo{author}{Casanova, J.}
\newblock \bibinfo{title}{Quantum simulation of dissipative processes without
  reservoir engineering}.
\newblock \emph{\bibinfo{journal}{Scientific Reports}}
  \textbf{\bibinfo{volume}{5}} (\bibinfo{year}{2015}).
\newblock \urlprefix\url{https://doi.org/10.1038/srep09981}.

\bibitem{Luchnikov2019}
\bibinfo{author}{Luchnikov, I.}, \bibinfo{author}{Vintskevich, S.},
  \bibinfo{author}{Ouerdane, H.} \& \bibinfo{author}{Filippov, S.}
\newblock \bibinfo{title}{Simulation complexity of open quantum dynamics:
  Connection with tensor networks}.
\newblock \emph{\bibinfo{journal}{Physical Review Letters}}
  \textbf{\bibinfo{volume}{122}} (\bibinfo{year}{2019}).
\newblock \urlprefix\url{https://doi.org/10.1103/physrevlett.122.160401}.

\bibitem{Loh1990}
\bibinfo{author}{Loh, E.~Y.} \emph{et~al.}
\newblock \bibinfo{title}{Sign problem in the numerical simulation of
  many-electron systems}.
\newblock \emph{\bibinfo{journal}{Physical Review B}}
  \textbf{\bibinfo{volume}{41}}, \bibinfo{pages}{9301--9307}
  (\bibinfo{year}{1990}).
\newblock \urlprefix\url{https://doi.org/10.1103/physrevb.41.9301}.

\bibitem{Prosen2007}
\bibinfo{author}{Prosen, T.} \& \bibinfo{author}{{\v{Z}}nidari{\v{c}}, M.}
\newblock \bibinfo{title}{Is the efficiency of classical simulations of quantum
  dynamics related to integrability?}
\newblock \emph{\bibinfo{journal}{Physical Review E}}
  \textbf{\bibinfo{volume}{75}} (\bibinfo{year}{2007}).
\newblock \urlprefix\url{https://doi.org/10.1103/physreve.75.015202}.

\bibitem{Breuer2007}
\bibinfo{author}{Breuer, H.-P.} \& \bibinfo{author}{Petruccione, F.}
\newblock \emph{\bibinfo{title}{The Theory of Open Quantum Systems}}
  (\bibinfo{publisher}{Oxford University Press}, \bibinfo{year}{2007}).
\newblock
  \urlprefix\url{https://doi.org/10.1093/acprof:oso/9780199213900.001.0001}.

\bibitem{Pfau2020}
\bibinfo{author}{Pfau, D.}, \bibinfo{author}{Spencer, J.~S.},
  \bibinfo{author}{Matthews, A. G. D.~G.} \& \bibinfo{author}{Foulkes, W.
  M.~C.}
\newblock \bibinfo{title}{Ab initio solution of the many-electron
  schr\"{o}dinger equation with deep neural networks}.
\newblock \emph{\bibinfo{journal}{Physical Review Research}}
  \textbf{\bibinfo{volume}{2}} (\bibinfo{year}{2020}).
\newblock \urlprefix\url{https://doi.org/10.1103/physrevresearch.2.033429}.

\bibitem{Hermann2020}
\bibinfo{author}{Hermann, J.}, \bibinfo{author}{Sch\"{a}tzle, Z.} \&
  \bibinfo{author}{No{\'{e}}, F.}
\newblock \bibinfo{title}{Deep-neural-network solution of the electronic
  schr\"{o}dinger equation}.
\newblock \emph{\bibinfo{journal}{Nature Chemistry}}
  \textbf{\bibinfo{volume}{12}}, \bibinfo{pages}{891--897}
  (\bibinfo{year}{2020}).
\newblock \urlprefix\url{https://doi.org/10.1038/s41557-020-0544-y}.

\bibitem{Zhu_2018}
\bibinfo{author}{Zhu, Y.} \& \bibinfo{author}{Zabaras, N.}
\newblock \bibinfo{title}{Bayesian deep convolutional encoder–decoder
  networks for surrogate modeling and uncertainty quantification}.
\newblock \emph{\bibinfo{journal}{Journal of Computational Physics}}
  \textbf{\bibinfo{volume}{366}}, \bibinfo{pages}{415–447}
  (\bibinfo{year}{2018}).
\newblock \urlprefix\url{http://dx.doi.org/10.1016/j.jcp.2018.04.018}.

\bibitem{Bhatnagar_2019}
\bibinfo{author}{Bhatnagar, S.}, \bibinfo{author}{Afshar, Y.},
  \bibinfo{author}{Pan, S.}, \bibinfo{author}{Duraisamy, K.} \&
  \bibinfo{author}{Kaushik, S.}
\newblock \bibinfo{title}{Prediction of aerodynamic flow fields using
  convolutional neural networks}.
\newblock \emph{\bibinfo{journal}{Computational Mechanics}}
  \textbf{\bibinfo{volume}{64}}, \bibinfo{pages}{525–545}
  (\bibinfo{year}{2019}).
\newblock \urlprefix\url{http://dx.doi.org/10.1007/s00466-019-01740-0}.

\bibitem{smith2020eikonet}
\bibinfo{author}{Smith, J.~D.}, \bibinfo{author}{Azizzadenesheli, K.} \&
  \bibinfo{author}{Ross, Z.~E.}
\newblock \bibinfo{title}{Eikonet: Solving the eikonal equation with deep
  neural networks}.
\newblock \emph{\bibinfo{journal}{{IEEE} Trans. Geosci. Remote. Sens.}}
  \textbf{\bibinfo{volume}{59}}, \bibinfo{pages}{10685--10696}
  (\bibinfo{year}{2021}).
\newblock \urlprefix\url{https://doi.org/10.1109/TGRS.2020.3039165}.

\bibitem{RAISSI2019686}
\bibinfo{author}{Raissi, M.}, \bibinfo{author}{Perdikaris, P.} \&
  \bibinfo{author}{Karniadakis, G.}
\newblock \bibinfo{title}{Physics-informed neural networks: A deep learning
  framework for solving forward and inverse problems involving nonlinear
  partial differential equations}.
\newblock \emph{\bibinfo{journal}{Journal of Computational Physics}}
  \textbf{\bibinfo{volume}{378}}, \bibinfo{pages}{686--707}
  (\bibinfo{year}{2019}).
\newblock \urlprefix\url{https://doi.org/10.1016/j.jcp.2018.10.045}.

\bibitem{raissi2018deep}
\bibinfo{author}{Raissi, M.}
\newblock \bibinfo{title}{Deep hidden physics models: Deep learning of
  nonlinear partial differential equations}.
\newblock \emph{\bibinfo{journal}{Journal of Machine Learning Research}}
  \textbf{\bibinfo{volume}{19}}, \bibinfo{pages}{25:1--25:24}
  (\bibinfo{year}{2018}).
\newblock \urlprefix\url{http://jmlr.org/papers/v19/18-046.html}.

\bibitem{lu2020deeponet}
\bibinfo{author}{Lu, L.}, \bibinfo{author}{Jin, P.} \&
  \bibinfo{author}{Karniadakis, G.~E.}
\newblock \bibinfo{title}{{DeepONet}: {L}earning nonlinear operators for
  identifying differential equations based on the universal approximation
  theorem of operators}.
\newblock \emph{\bibinfo{journal}{arXiv}} \textbf{\bibinfo{volume}{1910.03193}}
  (\bibinfo{year}{2020}).
\newblock \urlprefix\url{https://arxiv.org/abs/1910.03193}.

\bibitem{li2020neural}
\bibinfo{author}{Li, Z.} \emph{et~al.}
\newblock \bibinfo{title}{Neural operator: Graph kernel network for partial
  differential equations}.
\newblock \emph{\bibinfo{journal}{arXiv}} \textbf{\bibinfo{volume}{2003.03485}}
  (\bibinfo{year}{2020}).
\newblock \urlprefix\url{https://arxiv.org/abs/2003.03485}.

\bibitem{sanchezgonzalez2020learning}
\bibinfo{author}{Sanchez{-}Gonzalez, A.} \emph{et~al.}
\newblock \bibinfo{title}{Learning to simulate complex physics with graph
  networks}.
\newblock In \emph{\bibinfo{booktitle}{Proceedings of the 37th International
  Conference on Machine Learning, {ICML} 2020, 13-18 July 2020, Virtual
  Event}}, vol. \bibinfo{volume}{119} of \emph{\bibinfo{series}{Proceedings of
  Machine Learning Research}}, \bibinfo{pages}{8459--8468}
  (\bibinfo{publisher}{{PMLR}}, \bibinfo{year}{2020}).
\newblock
  \urlprefix\url{http://proceedings.mlr.press/v119/sanchez-gonzalez20a.html}.

\bibitem{cho2014learning}
\bibinfo{author}{Cho, K.} \emph{et~al.}
\newblock \bibinfo{title}{Learning phrase representations using {RNN}
  encoder-decoder for statistical machine translation}.
\newblock In \bibinfo{editor}{Moschitti, A.}, \bibinfo{editor}{Pang, B.} \&
  \bibinfo{editor}{Daelemans, W.} (eds.) \emph{\bibinfo{booktitle}{Proceedings
  of the 2014 Conference on Empirical Methods in Natural Language Processing,
  {EMNLP} 2014, October 25-29, 2014, Doha, Qatar, {A} meeting of SIGDAT, a
  Special Interest Group of the {ACL}}}, \bibinfo{pages}{1724--1734}
  (\bibinfo{publisher}{{ACL}}, \bibinfo{year}{2014}).
\newblock \urlprefix\url{https://doi.org/10.3115/v1/d14-1179}.

\bibitem{loshchilov2019decoupled}
\bibinfo{author}{Loshchilov, I.} \& \bibinfo{author}{Hutter, F.}
\newblock \bibinfo{title}{Decoupled weight decay regularization}.
\newblock In \emph{\bibinfo{booktitle}{7th International Conference on Learning
  Representations, {ICLR} 2019, New Orleans, LA, USA, May 6-9, 2019}}
  (\bibinfo{publisher}{OpenReview.net}, \bibinfo{year}{2019}).
\newblock \urlprefix\url{https://openreview.net/forum?id=Bkg6RiCqY7}.

\bibitem{Gers1999}
\bibinfo{author}{Gers, F.~A.}, \bibinfo{author}{Schmidhuber, J.} \&
  \bibinfo{author}{Cummins, F.~A.}
\newblock \bibinfo{title}{Learning to forget: Continual prediction with
  {LSTM}}.
\newblock \emph{\bibinfo{journal}{Neural Comput.}}
  \textbf{\bibinfo{volume}{12}}, \bibinfo{pages}{2451--2471}
  (\bibinfo{year}{2000}).
\newblock \urlprefix\url{https://doi.org/10.1162/089976600300015015}.

\bibitem{Simonyan2013}
\bibinfo{author}{Simonyan, K.}, \bibinfo{author}{Vedaldi, A.} \&
  \bibinfo{author}{Zisserman, A.}
\newblock \bibinfo{title}{Deep inside convolutional networks: Visualising image
  classification models and saliency maps}.
\newblock In \bibinfo{editor}{Bengio, Y.} \& \bibinfo{editor}{LeCun, Y.} (eds.)
  \emph{\bibinfo{booktitle}{2nd International Conference on Learning
  Representations, {ICLR} 2014, Banff, AB, Canada, April 14-16, 2014, Workshop
  Track Proceedings}} (\bibinfo{year}{2014}).
\newblock \urlprefix\url{http://arxiv.org/abs/1312.6034}.

\bibitem{Sundararajan2017}
\bibinfo{author}{Sundararajan, M.}, \bibinfo{author}{Taly, A.} \&
  \bibinfo{author}{Yan, Q.}
\newblock \bibinfo{title}{Axiomatic attribution for deep networks}.
\newblock In \bibinfo{editor}{Precup, D.} \& \bibinfo{editor}{Teh, Y.~W.}
  (eds.) \emph{\bibinfo{booktitle}{Proceedings of the 34th International
  Conference on Machine Learning, {ICML} 2017, Sydney, NSW, Australia, 6-11
  August 2017}}, vol.~\bibinfo{volume}{70} of
  \emph{\bibinfo{series}{Proceedings of Machine Learning Research}},
  \bibinfo{pages}{3319--3328} (\bibinfo{publisher}{{PMLR}},
  \bibinfo{year}{2017}).
\newblock
  \urlprefix\url{http://proceedings.mlr.press/v70/sundararajan17a.html}.

\bibitem{Riemann1851}
\bibinfo{author}{Riemann, B.}
\newblock \bibinfo{title}{Grundlagen f\"ur eine allgemeine {T}heorie der
  {F}unktionen einer ver\"anderlichen komplexen {G}r\"osse (1851)}.
\newblock In \bibinfo{editor}{Weber, H.} (ed.)
  \emph{\bibinfo{booktitle}{Riemann's gesammelte math. {W}erke (in {G}erman)}},
  \bibinfo{pages}{3--48} (\bibinfo{publisher}{Dover}, \bibinfo{year}{1953}).

\bibitem{Zdeborov2017}
\bibinfo{author}{Zdeborov{\'{a}}, L.}
\newblock \bibinfo{title}{New tool in the box}.
\newblock \emph{\bibinfo{journal}{Nature Physics}}
  \textbf{\bibinfo{volume}{13}}, \bibinfo{pages}{420--421}
  (\bibinfo{year}{2017}).
\newblock \urlprefix\url{https://doi.org/10.1038/nphys4053}.

\bibitem{Zdeborov2020}
\bibinfo{author}{Zdeborov{\'{a}}, L.}
\newblock \bibinfo{title}{Understanding deep learning is also a job for
  physicists}.
\newblock \emph{\bibinfo{journal}{Nature Physics}}
  \textbf{\bibinfo{volume}{16}}, \bibinfo{pages}{602--604}
  (\bibinfo{year}{2020}).
\newblock \urlprefix\url{https://doi.org/10.1038/s41567-020-0929-2}.

\bibitem{Bapst2020}
\bibinfo{author}{Bapst, V.} \emph{et~al.}
\newblock \bibinfo{title}{Unveiling the predictive power of static structure in
  glassy systems}.
\newblock \emph{\bibinfo{journal}{Nature Physics}}
  \textbf{\bibinfo{volume}{16}}, \bibinfo{pages}{448--454}
  (\bibinfo{year}{2020}).
\newblock \urlprefix\url{https://doi.org/10.1038/s41567-020-0842-8}.

\bibitem{Lee2022}
\bibinfo{author}{Lee, K.}, \bibinfo{author}{Ayyasamy, M.~V.},
  \bibinfo{author}{Delsa, P.}, \bibinfo{author}{Hartnett, T.~Q.} \&
  \bibinfo{author}{Balachandran, P.~V.}
\newblock \bibinfo{title}{Phase classification of multi-principal element
  alloys via interpretable machine learning}.
\newblock \emph{\bibinfo{journal}{npj Computational Materials}}
  \textbf{\bibinfo{volume}{8}} (\bibinfo{year}{2022}).
\newblock \urlprefix\url{https://doi.org/10.1038/s41524-022-00704-y}.

\bibitem{Balestriero2021}
\bibinfo{author}{Balestriero, R.}, \bibinfo{author}{Pesenti, J.} \&
  \bibinfo{author}{LeCun, Y.}
\newblock \bibinfo{title}{Learning in high dimension always amounts to
  extrapolation}.
\newblock \emph{\bibinfo{journal}{arXiv}} \textbf{\bibinfo{volume}{2110.09485}}
  (\bibinfo{year}{2021}).
\newblock \urlprefix\url{https://arxiv.org/abs/2110.09485}.

\bibitem{Wald2021}
\bibinfo{author}{Wald, Y.}, \bibinfo{author}{Feder, A.},
  \bibinfo{author}{Greenfeld, D.} \& \bibinfo{author}{Shalit, U.}
\newblock \bibinfo{title}{On calibration and out-of-domain generalization}.
\newblock \emph{\bibinfo{journal}{arXiv}} \textbf{\bibinfo{volume}{2102.10395}}
  (\bibinfo{year}{2021}).
\newblock \urlprefix\url{https://arxiv.org/abs/2102.10395}.

\bibitem{Wang2021}
\bibinfo{author}{Wang, J.}, \bibinfo{author}{Lan, C.}, \bibinfo{author}{Liu,
  C.}, \bibinfo{author}{Ouyang, Y.} \& \bibinfo{author}{Qin, T.}
\newblock \bibinfo{title}{Generalizing to unseen domains: {A} survey on domain
  generalization}.
\newblock In \bibinfo{editor}{Zhou, Z.} (ed.)
  \emph{\bibinfo{booktitle}{Proceedings of the Thirtieth International Joint
  Conference on Artificial Intelligence, {IJCAI} 2021, Virtual Event /
  Montreal, Canada, 19-27 August 2021}}, \bibinfo{pages}{4627--4635}
  (\bibinfo{publisher}{ijcai.org}, \bibinfo{year}{2021}).
\newblock \urlprefix\url{https://doi.org/10.24963/ijcai.2021/628}.

\bibitem{Pang2021}
\bibinfo{author}{Pang, G.}, \bibinfo{author}{Shen, C.}, \bibinfo{author}{Cao,
  L.} \& \bibinfo{author}{van~den Hengel, A.}
\newblock \bibinfo{title}{Deep learning for anomaly detection: {A} review}.
\newblock \emph{\bibinfo{journal}{{ACM} Comput. Surv.}}
  \textbf{\bibinfo{volume}{54}}, \bibinfo{pages}{38:1--38:38}
  (\bibinfo{year}{2021}).
\newblock \urlprefix\url{https://doi.org/10.1145/3439950}.

\bibitem{Yang2021}
\bibinfo{author}{Yang, J.}, \bibinfo{author}{Zhou, K.}, \bibinfo{author}{Li,
  Y.} \& \bibinfo{author}{Liu, Z.}
\newblock \bibinfo{title}{Generalized out-of-distribution detection: {A}
  survey}.
\newblock \emph{\bibinfo{journal}{arXiv}} \textbf{\bibinfo{volume}{2110.11334}}
  (\bibinfo{year}{2021}).
\newblock \urlprefix\url{https://arxiv.org/abs/2110.11334}.

\bibitem{Liu2020}
\bibinfo{author}{Liu, W.}, \bibinfo{author}{Wang, X.}, \bibinfo{author}{Owens,
  J.~D.} \& \bibinfo{author}{Li, Y.}
\newblock \bibinfo{title}{Energy-based out-of-distribution detection}.
\newblock In \bibinfo{editor}{Larochelle, H.}, \bibinfo{editor}{Ranzato, M.},
  \bibinfo{editor}{Hadsell, R.}, \bibinfo{editor}{Balcan, M.} \&
  \bibinfo{editor}{Lin, H.} (eds.) \emph{\bibinfo{booktitle}{Advances in Neural
  Information Processing Systems 33: Annual Conference on Neural Information
  Processing Systems 2020, NeurIPS 2020, December 6-12, 2020, virtual}}
  (\bibinfo{year}{2020}).
\newblock
  \urlprefix\url{https://proceedings.neurips.cc/paper/2020/hash/f5496252609c43eb8a3d147ab9b9c006-Abstract.html}.

\bibitem{NAKANO1994181}
\bibinfo{author}{Nakano, A.}, \bibinfo{author}{Vashishta, P.} \&
  \bibinfo{author}{Kalia, R.~K.}
\newblock \bibinfo{title}{Massively parallel algorithms for computational
  nanoelectronics based on quantum molecular dynamics}.
\newblock \emph{\bibinfo{journal}{Computer Physics Communications}}
  \textbf{\bibinfo{volume}{83}}, \bibinfo{pages}{181--196}
  (\bibinfo{year}{1994}).
\newblock
  \urlprefix\url{https://www.sciencedirect.com/science/article/pii/0010465594900477}.

\bibitem{Bengio1994}
\bibinfo{author}{Bengio, Y.}, \bibinfo{author}{Simard, P.~Y.} \&
  \bibinfo{author}{Frasconi, P.}
\newblock \bibinfo{title}{Learning long-term dependencies with gradient descent
  is difficult}.
\newblock \emph{\bibinfo{journal}{{IEEE} Trans. Neural Networks}}
  \textbf{\bibinfo{volume}{5}}, \bibinfo{pages}{157--166}
  (\bibinfo{year}{1994}).
\newblock \urlprefix\url{https://doi.org/10.1109/72.279181}.

\bibitem{pascanu2013difficulty}
\bibinfo{author}{Pascanu, R.}, \bibinfo{author}{Mikolov, T.} \&
  \bibinfo{author}{Bengio, Y.}
\newblock \bibinfo{title}{On the difficulty of training recurrent neural
  networks}.
\newblock In \emph{\bibinfo{booktitle}{Proceedings of the 30th International
  Conference on Machine Learning, {ICML} 2013, Atlanta, GA, USA, 16-21 June
  2013}}, vol.~\bibinfo{volume}{28} of \emph{\bibinfo{series}{{JMLR} Workshop
  and Conference Proceedings}}, \bibinfo{pages}{1310--1318}
  (\bibinfo{publisher}{JMLR.org}, \bibinfo{year}{2013}).
\newblock \urlprefix\url{http://proceedings.mlr.press/v28/pascanu13.html}.

\bibitem{kingma2017adam}
\bibinfo{author}{Kingma, D.~P.} \& \bibinfo{author}{Ba, J.}
\newblock \bibinfo{title}{Adam: {A} method for stochastic optimization}.
\newblock In \bibinfo{editor}{Bengio, Y.} \& \bibinfo{editor}{LeCun, Y.} (eds.)
  \emph{\bibinfo{booktitle}{3rd International Conference on Learning
  Representations, {ICLR} 2015, San Diego, CA, USA, May 7-9, 2015, Conference
  Track Proceedings}} (\bibinfo{year}{2015}).
\newblock \urlprefix\url{http://arxiv.org/abs/1412.6980}.

\end{thebibliography}

%%%%%%%%%%%%%%%%%%%%%%%%%%%%%%%%%%%%%%%%%%%%%%%%%%%%%%%%%%%%%%%%%:~
\cleardoublepage

\section*{Methods}

\subsection*{Problem Formulation} We consider the one-dimensional time-dependent Schrödinger equation in atomic units,
    \begin{equation}
        i \frac{\partial}{\partial t} \psi(x, t)= \mathcal{H} \psi(x, t),
    \end{equation}
    where $x \in [0, L_x), t \in [0, T)$, and the Hamiltonian operator $\mathcal{H}$ is defined as
    \begin{equation}
        \mathcal{H} = T + V = -\frac{1}{2} \frac{\partial^{2}}{\partial x^{2}}+V(x),
    \end{equation}
    with $T$ and $V$ being the kinetic and potential energy operators, respectively.
    
    We represent the quantum waves as complex-valued functions, $\psi(x, t) \in \mathbb{C}$, on a $N_x \times N_t$ mesh grid, and
    we impose periodic boundary conditions: $\psi(x+L_x, t)=\psi(x, t)$, where $L_x$ is the size of our 1-dimensional environment.
    
    We use an equal mesh spacing. With $\Delta x = L_x / N_x$ and $\Delta t = T / N_t$, the spatial and temporal coordinates become $x_i=i \Delta x$ and $t_j=j \Delta t$, 
    with $i \in \{0, 1, \ldots, N_x - 1 \}$ and $j \in \{0, 1, \ldots, N_t - 1 \}$, respectively.
    For brevity, we denote the discrete wave function values as $\psi_{i}^j=\psi(x_i, t_j)$, and the discrete potential steps as $v_i = V(x_i)$.

    Given the discretizations described above, we consider the neural network-based emulator as a parameterized map from an input, constructed from the $H$ consecutive time steps, $\{t_{j}, t_{t+1}, \ldots, t_{j+H-1} \}$, to the output that relates to the next time step, $t_{j+H}$.
    To make the input and the output of the neural network independent from the size of the system, we construct a slices (we call them \emph{windows} through the text) of a fixed width $W$.
    Namely, a portion of the input representing the values of the quantum wave at the time step $t_j$, can be written as
    \begin{widetext}
    \begin{equation}
        \boldsymbol{\omega}_i^j = \left(\psi_{i-\lfloor W/2 \rfloor}^j, \psi_{i-\lfloor W/2 \rfloor+1}^j, \ldots, \psi_{i}^j, \ldots, \psi_{i+\lfloor (W-1)/2 \rfloor-1}^j, \psi_{i+\lfloor (W-1)/2 \rfloor}^j \right).
    \label{window def}
    \end{equation}
    Similarly, a portion of the input representing the potential landscape at the same time step, can be written as
    \begin{equation}
        \boldsymbol{\nu}_i= \left( v_{i-\lfloor W/2 \rfloor}, v_{i-\lfloor W/2 \rfloor+1}, \ldots, v_{i}, \ldots, v_{i+\lfloor (W-1)/2 \rfloor-1}, v_{i+\lfloor (W-1)/2 \rfloor} \right).
    \label{pot window def}
    \end{equation}
    \end{widetext}

    As shown in Fig.~\ref{fig:nn}, the input of raw simulation data can be decomposed into a series of windows. The proposed neural network-based emulator maps each input data window to an output data window, that can be subsequently re-combined to form a complete time step. More specifically, we look for a network to learn the mapping
    \begin{equation}
        \label{eq:mapping}
        f_{\boldsymbol{\Theta}}: 
        \begin{Bmatrix}
        \boldsymbol{\omega}_i^j,\boldsymbol{\omega}_i^{j+1},\ldots,\boldsymbol{\omega}_i^{j+H-1},\boldsymbol{\nu}_i
        \end{Bmatrix} \rightarrow 
        \begin{Bmatrix}
        \boldsymbol{\omega}_i^{j+H}
        \end{Bmatrix}.
    \end{equation}
    During the training, we try to find parameters $\boldsymbol{\Theta}$ of the map $f_{\boldsymbol{\Theta}}$, that minimize the training objective function (a loss function $\mathcal{L}$) on each window output, defined as
    \begin{equation}
    \mathcal{L} = \mbox{MSE} _i ^j=\frac{1}{W}	\left \| \boldsymbol{\hat{\omega}}_i^{j+H} - \boldsymbol{\omega}_i^{j+H} \right \|^{2},
    \end{equation}
    where $\boldsymbol{\hat{\omega}}_i^{j+H}$ denotes the predicted values of the quantum wave at the time step $t_j$ in the considered window, and $\boldsymbol{{\omega}}_i^{j+H}$ denotes the ground truth values.
    
\subsection*{Prediction using the window based scheme}
\label{sec:met-pred}

    To make the emulator independent of the environment size, we predict only the local evolution of the wave packet $\boldsymbol{\omega}_i^j$ within a given window (cf.~Eq.~\ref{window def}). However, to recreate the entire wave function $\psi(x, t_j)$ for a given time step $j$, we need to assemble the individual predictions from different, overlapping windows.
    Let us denote the $\psi_m^j$ evaluated from window $\boldsymbol{\omega}_{m+k}^{j}$ as $\psi_m^j\left(\boldsymbol{\omega}_{m+k}^{j}\right)$.
    We average those predictions by giving a larger weight to those predictions for which the grid-point $x_m$ is located closer to the center of the window.
    For the weight, we use the Gaussian modulation,
    \begin{equation}
    \langle \psi_{m}^{j} \rangle = \sum_{k=-\lfloor (W-1) / 2\rfloor}^{\lfloor W / 2\rfloor} A\, \exp\!{\left(-\frac{k^{2}}{2 \delta^{2}}\right)} \psi_{i}^{j}\left(\boldsymbol{\omega}_{m+k}^{j}\right),
    \end{equation}
    where $A=1 / \sum_{k=-\lfloor (W-1) / 2\rfloor}^{\lfloor W / 2\rfloor} \exp\!{\left(-k^{2} \middle/ 2 \delta^{2}\right)}$ is the normalization constant, and $\delta$ denote the Gaussian averaging spread.
    
    A complete prediction of the next time-step is a collection of all predicted points $[\langle \psi_{0}^{j} \rangle, \langle \psi_{1}^{j} \rangle, \ldots, \langle \psi_{N_x}^{j} \rangle]$.
    Fortunately, all the intermediate predictions $\psi_m^j\left(\boldsymbol{\omega}_{m+k}^{j}\right)$ can be obtained independently, therefore the entire algorithm is easily parallelizable (e.g., by performing a batch inference on a modern GPU).

\subsection*{Predicting multiple steps of evolution}

    We predict the next time-step of the system evolution in a recurrent manner, i.e., the predictions of the previous time steps are used to form the input of the current time-step.
    By iteratively predicting next time-steps, we can obtain a sequence of snapshots, portraying the wave function evolution, of an arbitrary length.
    
    As an initial condition, the trained models are fed in with first four time steps of the system evolution.

\subsection*{Data generation and processing}
\label{sec:method_data}

    Our data generation implements a two-step procedure. First, we generate raw simulation data that represents the state of the system. Then we slice them into smaller windows to feed into the neural network.

\paragraph*{Raw simulation.}

    \begin{table*}
    \centering
    \setlength{\tabcolsep}{12pt}
    \caption{\textbf{Raw Simulation Parameters.} Parameters describing the spatial and temporal discretization of the wave function evolution ($N_x$, $\Delta t$), the environment together with a centrally-located rectangular potential barrier ($L_x$, $H_b$, $W_b$), and the initial state of the wave packet in the time~$t_0$ ($X_0$, $S_0$, $E_0$).}
    \begin{tabularx}{\linewidth}{cX}  %{ c p{0.82\linewidth} }
     \hline
     \hline
     \(N_x\), \(L_x\) & The wave function is discretized on a regular mesh of size  \(\Delta x  = L_x/N_x\) where $N_x$ is the number of mesh points and $L_x$ is the length of the simulation box. \\ 
     \hline
     \(\Delta t\) & Time discretization unit. The time between each discrete time-step. \\ 
     \hline
     \(X_0\), \(S_0\), \(E_0\) & Center-of-mass, spread and energy of initial Gaussian-modulated wave packet, respectively. \\ 
     \hline
     \(H_b\), \(W_b\) & Rectangular barrier height and width centered at the simulation box, respectively.\\
     \hline
    \hline
    \end{tabularx}
    \label{tab:sim-para}
    \end{table*}

    We generate raw simulation data of wave propagation discretized on a regular mesh of size $\Delta x$ using a space splitting method \cite{NAKANO1994181}. We simulate the wave propagation over a rectangular potential barrier centered at the simulation box. The important parameters controlling the simulations are listed in Tab.~\ref{tab:sim-para}.
    Specifically, we set \(L_x=100 \, a.u.\), \(N_x=1024\), and \(W_b=7.0\, a.u\). In $t=0$, the wave packet has a Gaussian modulation,
    \begin{equation}
    \psi_{t=0}(x)=C \exp{\left(-\frac{(x-X_0)^2}{4S_0^2}\right)}\, \exp{\left(i\sqrt{2E_0}x\right)}.
    \end{equation}
    The simulations are running with an internal-time step of $\Delta t_{int}=0.0005 \, a.u$.
    We run the simulation for $100.000$ steps, while saving snapshots of the simulation by every 200 steps (thus $\Delta t=0.1 \, a.u$).
    As a result, we get $(N_t=500) \times (N_x=1024)$ records of the wave function evolution. In case of a free propagation, each pixel is represented by two values, the real and the imaginary part of the wave function. In the simulations including a potential barrier, we add a third channel to each pixel, to encode values of the potential at each point.  

\paragraph*{Data windowing.}

    To make the input of the neural network independent of the environment size, 
    we slice the recorded raw simulation data into windows of size $(H+1) \times W \times C$. Specifically, $W=23$ represents the spatial window size, $C=2$ or $3$ denotes the number of information channels, $H+1=5$ stands for the number of consequent time steps. The first $H$ time steps are used to construct the training data set. The last time step in each simulation is used to construct the training target.
    
\paragraph*{Data processing.}

    We adopted a few techniques to efficiently process the training data.
    
    \begin{enumerate}[leftmargin=*,itemsep=2pt,topsep=2pt]
    \item To reduce the correlations between training samples, we apply spatial and temporal sampling to intentionally skip some windows. We set the spatial sampling ratio to  0.1 and the temporal ratio to  0.9.
    \item We assign a higher possibility to data windows overlapped with the central potential barrier. This helps us balance the ``hard'' and ``easy'' cases in the training set.
    \item We apply the periodic boundary condition.
    \item We renormalize the values in the channels. Specificly, the potential values are rescaled to [0, 1] to improve the training stability (e.g., to avoid the \emph{exploding gradient} problem\cite{Bengio1994, pascanu2013difficulty}).
    \end{enumerate}

\subsection*{Description of the neural network architectures}
\label{sec:architectures}
    The input of the neural network has dimension $4 \times 23 \times 3$ (four time steps, window size $23$, three channels). First the input tensor is reshaped into a $4 \times 69$ tensor, and then connected to a time-distributed dense layer (a dense layer that is applied independently to each time step) with the ReLU activation functions. We treat the size of the dense layer, $K$, as a tunable hyper-parameter. We pass the resulting $4 \times K$ tensor to a gated recurrent unit, GRU \cite{cho2014learning} with the ReLU activation as well. The output of that step is a single tensor of size $K$. We connect it to the output layer of size $46$, with a linear activation (see again the schematic representation of the network in Fig.~\ref{fig:nn}{\bf a}).
    
    In addition to the architecture described above, we used the following three benchmark network architectures:
    \begin{enumerate}[leftmargin=*]
        \item \emph{Linear model}. For this model, we used a simplified input. Instead of four time-steps, the linear model takes a single time-step (thus, the input tensor shape is $1 \times 23 \times 3$). We connect the input directly to the output layer of size $46$. There is no hidden layers. All activations are linear.
        \item \emph{Fully dense model}. For the input we take the standard four time-steps (represented by a \(4 \times 23 \times 3\) tensor). First, the input is reshaped into a $4 \times 69$ tensor. Next, each time step of the input goes through a time-distributed dense layer of size $K$ with the ReLU activation. The output of the time-distributed layer is subsequently flattened and connected to a regular dense layers of size $K$, also with the ReLU activation. The resulting tensor is finally connected to the output layer of size $46$ with a linear activation.
        \item \emph{Convolutional model}. We use the same input as above, four time-steps represented as \(4 \times 23 \times 3\) tensors. We apply 1-dimensional convolution with a filter size of $4$ and with the ReLU activation. The purpose of that layer is to mix the original channels (encoding the real and imaginary part of the wave packet as well as the values of the potential) from different temporal points. The resulting tensor has shape $1 \times 23 \times F$, where $F$ is the number of the convolution filters. We took $F = \lfloor K/4 \rfloor$. The resulting tensor is subsequently flattened and connected to a hidden layer with $K$ neurons, and with the ReLU activation. The last layer is the output of size $46$ with a linear activation.
    \end{enumerate}
    The outputs of all the models have sizes of 46. In the last step, the outputs are reshaped to form $23 \times 2$ tensors, with the second dimension encoding the real and imaginary part of the wave function.

\subsection*{Training details}

   We have trained the neural networks using AdamW \cite{loshchilov2019decoupled} optimizer with the mean squared error (MSE) loss function. We used a linearly-decayed learning rate with a warm-up. We included more information regarding the training procedure in the Supplementary Information.
    
\subsection*{Model evaluation}
    To evaluate the model's performance, we compare the predicted evolution of the system with the ground truth, using the following two metrics:
    \begin{enumerate}
        \item Mean absolute error (MAE), calculated per each time step,
            \begin{equation}
                \overline{|\epsilon|} = \sum_{i=1}^{N_x} \frac{{|\hat{\psi_i}-\psi_i|}}{N_x},
            \end{equation}
            where $N_x$ is the number of spatial points, while $\hat{\psi_i}$ and $\psi_i$ are the ground truth and the predicted wave function value, respectively, evaluated in the i-th discrete position, $x_i$. MAE remains at low values for a good model. However, a consistent low MAE is not sufficient to determine a good model, as a model that constantly outputs zero or near-zero values might result in a relatively low MAE in some cases (e.g., when the ground truth is a wave packet concentrated only in a small volume of a larger space). This limitation of MAE, leads us to introduce another metric, namely, the normalized correlation.
        \item Normalized correlation per time step:
            \begin{equation}
                \mathcal{C}= \frac{\sum_{i=1}^{N_x} {\hat{\psi_i}^* \psi_i}}{|\boldsymbol{\hat{\psi}}||\boldsymbol{\psi}|},
            \end{equation}
            where, $N_x$, $\hat{\psi_i}$ and $\psi_i$ are defined in the same way as above. The symbol $^*$ represents the complex conjugate. Normalized correlation treats the predicted and true wave functions as two vectors, i.e. $\boldsymbol{\hat{\psi}}$ and $\boldsymbol{\psi}$. To this end, normalized correlation can be understood as the angular similarity between two wave function vectors. 
    \end{enumerate}
    For an overall performance evaluation, we calculate the average over a number of consequitive time steps, denoted by $\langle\overline{|\epsilon|}\rangle$ and $\langle \mathcal{C}\rangle$ respectively.

%%%%%%%%%%%%%%%%%%%%%%%%%%%%%%%%%%%%%%%%%%%%%%%%%%%%%%%%%%%%%%%%%:~

\section*{Data Availability}

    A sample data can be generated using scripts, available in our GitHub repository, \url{https://github.com/yaoyu-33/quantum_dynamics_machine_learning}.

    The full training data set as well as all validation and test cases are available from the corresponding author upon reasonable request.

%%%%%%%%%%%%%%%%%%%%%%%%%%%%%%%%%%%%%%%%%%%%%%%%%%%%%%%%%%%%%%%%%:~

\section*{Code Availability}

    Source code for training and evaluating the machine learning models is available at \url{https://github.com/yaoyu-33/quantum_dynamics_machine_learning}.

%%%%%%%%%%%%%%%%%%%%%%%%%%%%%%%%%%%%%%%%%%%%%%%%%%%%%%%%%%%%%%%%%:~

%\section*{References}% Un-comment me after we copy here the output of the .bbl file. 
%\bibliographystyle{naturemag}  % Nature Bibtex Style
%\def\bibsection{\section*{\refname}} 
%\bibliography{bibliography.bib}

%%%%%%%%%%%%%%%%%%%%%%%%%%%%%%%%%%%%%%%%%%%%%%%%%%%%%%%%%%%%%%%%%:~

\section*{Acknowledgments}
\label{sec:Acknowledgments}
% include Oleg, Aiichiro

    The authors acknowledge the Center for Advanced Research Computing (CARC) at the University of Southern California for providing computing resources that have contributed to the research results reported within this publication. URL: \url{https://carc.usc.edu}.

    We would like to thank Prof.\ Oleg V. Prezhdo and Prof.\ Aiichiro Nakano for helpful discussions during the early stages of the project.

%%%%%%%%%%%%%%%%%%%%%%%%%%%%%%%%%%%%%%%%%%%%%%%%%%%%%%%%%%%%%%%%%:~

\section*{Author Contributions}
\label{sec:Contribution}

    The study was planned by M.Ab.\ and S.H. The manuscript was prepared by Y.Y., C.C, S.H., and M.Ab. The data set construction was done by Y.Y. and C.C. The machine learning studies were performed by both Y.Y.\ and C.C. Result validation and code optimization was performed by M.Ag., D.K., and M.Ab. All authors discussed the results, wrote, and commented on the manuscript.

%%%%%%%%%%%%%%%%%%%%%%%%%%%%%%%%%%%%%%%%%%%%%%%%%%%%%%%%%%%%%%%%%:~

\section*{Competing Interests}

    The authors declare no competing interests.

%%%%%%%%%%%%%%%%%%%%%%%%%%%%%%%%%%%%%%%%%%%%%%%%%%%%%%%%%%%%%%%%%:~

\clearpage
\onecolumngrid
\part*{Supplementary Information}
\renewcommand\thefigure{S\arabic{figure}}    
\setcounter{figure}{0}

\section*{Choice of Examples}

    One of the goals of our work was to demonstrate the following possibility: \emph{it is feasible to emulate certain physical processes using machine learning models, that had been trained solely on some restricted, simple examples}.
    In other words, our aim was to address the particular challenge of training machine learning models in the reality, where generating training examples is difficult or expensive with exception of some specific discrete cases.
    To deliver a clear proof of concept, we have deliberately chosen a simple and well known problem: \emph{one-dimensional quantum dynamics}.
    To train the model, we have used only examples concerning a single, rectangular potential barrier and a single wave packet with Gaussian modulation. To quantify the extend to which the model generalize, we have designed a number of test cases consist of various potential barrier shapes and compositions.
    %For this task, we represented the input data using a window-based scheme as described in the main text. This idea was inspired by the locality of dynamics governed by partial differential equations.
    
    To that extend, we have demonstrated that it is possible to emulate the dynamics of quantum mechanical systems, without explicit knowledge of the Schr\"odinger equation or any physical laws in general. 
    In contrast to previous studies, we focused on eliciting what neural networks need for rendering faithful emulations and what the neural network learns during the training. 
    
    \section*{Details on the Training Data Set}
    
    We generated training examples by evolving a single Gaussian-modulated wave packet. The wave packet can be described with three parameters $X_0$, $S_0$, and $E_0$, which denote the center-of-mass position, the spread, and the energy, respectively. 
    
    During our training, we have considered two training regimes. The first one includes examples of free propagation (no potential barrier), while the second is concerned with propagation through an environment with a single rectangular barrier. If present, we constrain the barrier to be located at the center of the simulation box. Thus, the potential can be defined by two quantities, $W_b$ and $H_b$, which are the height and width of the barrier, respectively.
    
    Below, we show the values of parameters used when generating the training examples.
    
\paragraph*{Emulation of freely dispersing wave packets:}
    \begin{itemize}[itemsep=1pt,topsep=5pt]
        \item[] $X_0=10.0,\ 40.0,\ 70.0\,\ a.u.$
        \item[] $S_0=1.0,\ 1.5,\ 2.0,\ 2.5,\ 3.0,\ 3.5,\ 4.0\,\ a.u.$
        \item[] $E_0=1.0,\ 2.0,\ 3.0,\ 4.0,\ 5.0,\ 6.0,\ 7.0,\ 8.0,\ 9.0\,\ a.u.$
    \end{itemize}

\paragraph*{Quantum wave emulations with potential barrier being present:}
    \begin{itemize}[itemsep=1pt,topsep=5pt]
        \item[] $X_0=10.0,\ 40.0,\ 70.0\,\ a.u.$
        \item[] $S_0=1.0,\ 1.5,\ 2.0,\ 2.5,\ 3.0,\ 3.5,\ 4.0\,\ a.u.$
        \item[] $E_0=1.0,\ 2.0,\ 3.0,\ 4.0,\ 5.0,\ 6.0,\ 7.0,\ 8.0,\ 9.0\,\ a.u.$
        \item[] $H_b=1.0,\ 2.0,\ 3.0,\ 4.0,\ 5.0,\ 6.0,\ 7.0,\ 8.0,\ 9.0,\ 10.0,\ 11.0,\ 12.0,\ 13.0,\ 14.0\,\ a.u.$
        \item[] $W_b=7.0\,\ a.u.$
    \end{itemize}    
    
\begin{figure*}
        \begin{center}
            \includegraphics[width=\textwidth]{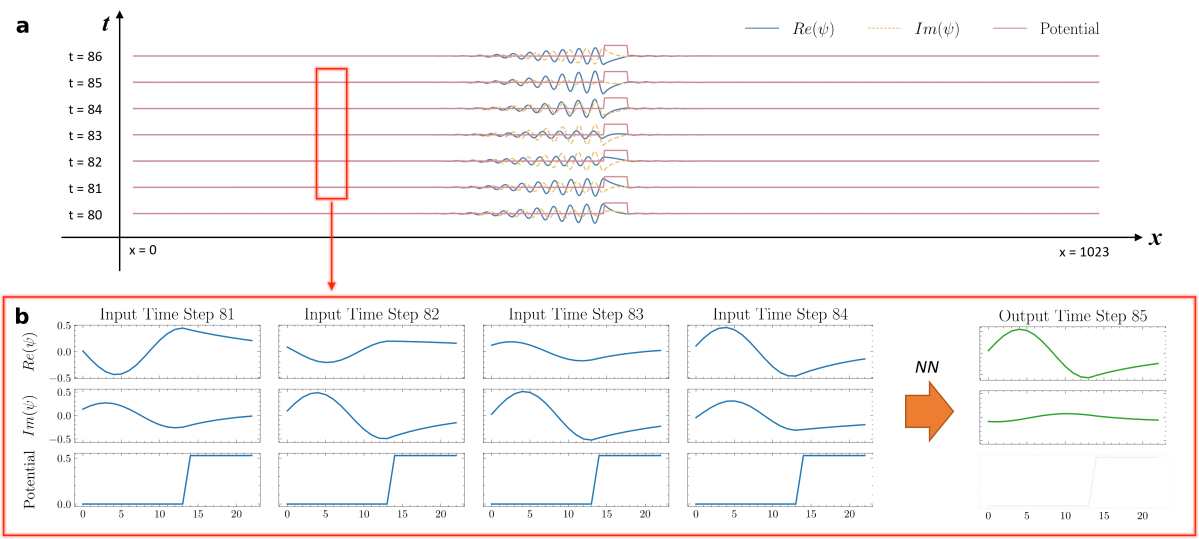}
        \end{center}
         \caption{\textbf{Structure of the input.} {\bf a}, The data input consists of a time series of one-dimensional plots, depicting the real and the imaginary part of the wave function as well as the values of the potential barrier. {\bf b}, All input information can be represented as a tensor, with three channels (similar to RGB channels when representing visual data). By selecting \emph{windows} of a fixed width, we can work with the data input of arbitrary length. Note, that the target has just two channels (not three), since we do not have to predict the potential landscape, only the wave function.} 
        \label{fig:time_series_input}
        \end{figure*}

    \vspace{6pt}
    In Figure~\ref{fig:time_series_input}, we show in detail the input structure used for training the neural network-based emulator. The red rectangular indicates the cutting window, that produces slices of data from the recorded simulations. Those slices are than reassembled into a time series, that consist of the input of a neural network.
    
\begin{figure*}
        \begin{center}
            \includegraphics[width=\textwidth]{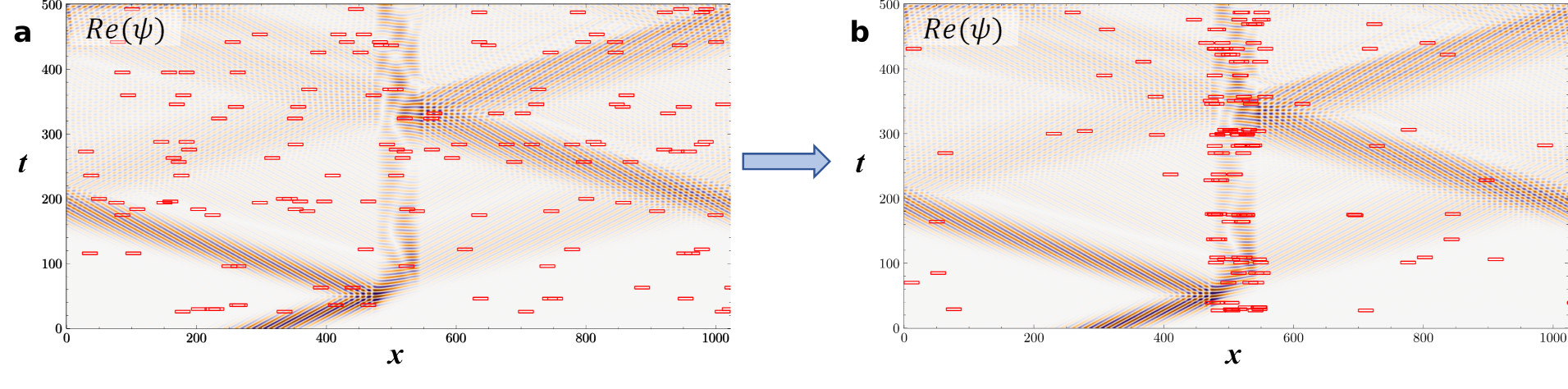}
        \end{center}
         \caption{\textbf{Input Sampling.} By sampling from all possible inputs, we create a training curriculum. We focus on balancing examples that illustrate different physical phenomena. In this case, we put an extra focus on the wave propagation near the rectangular potential barrier.} 
        \label{fig:input_sampling}
        \end{figure*}

    In Figure~\ref{fig:input_sampling}, we present the purpose of the input sampling. We assign a higher possibility to data windows that overlaps with the central potential barrier. The motivation is, that these cases are inherently harder to learn than examples depicting a free propagation. By tuning the sampling probability, we can easily control the ratio between the ``no-potential'' and ``with-potential'' windows, which helps to balance the training data set.

\section*{Hand-Designed Test Data Sets}
\label{sec:full_test}

\begin{figure*}
        \begin{center}
            \includegraphics[width=\textwidth]{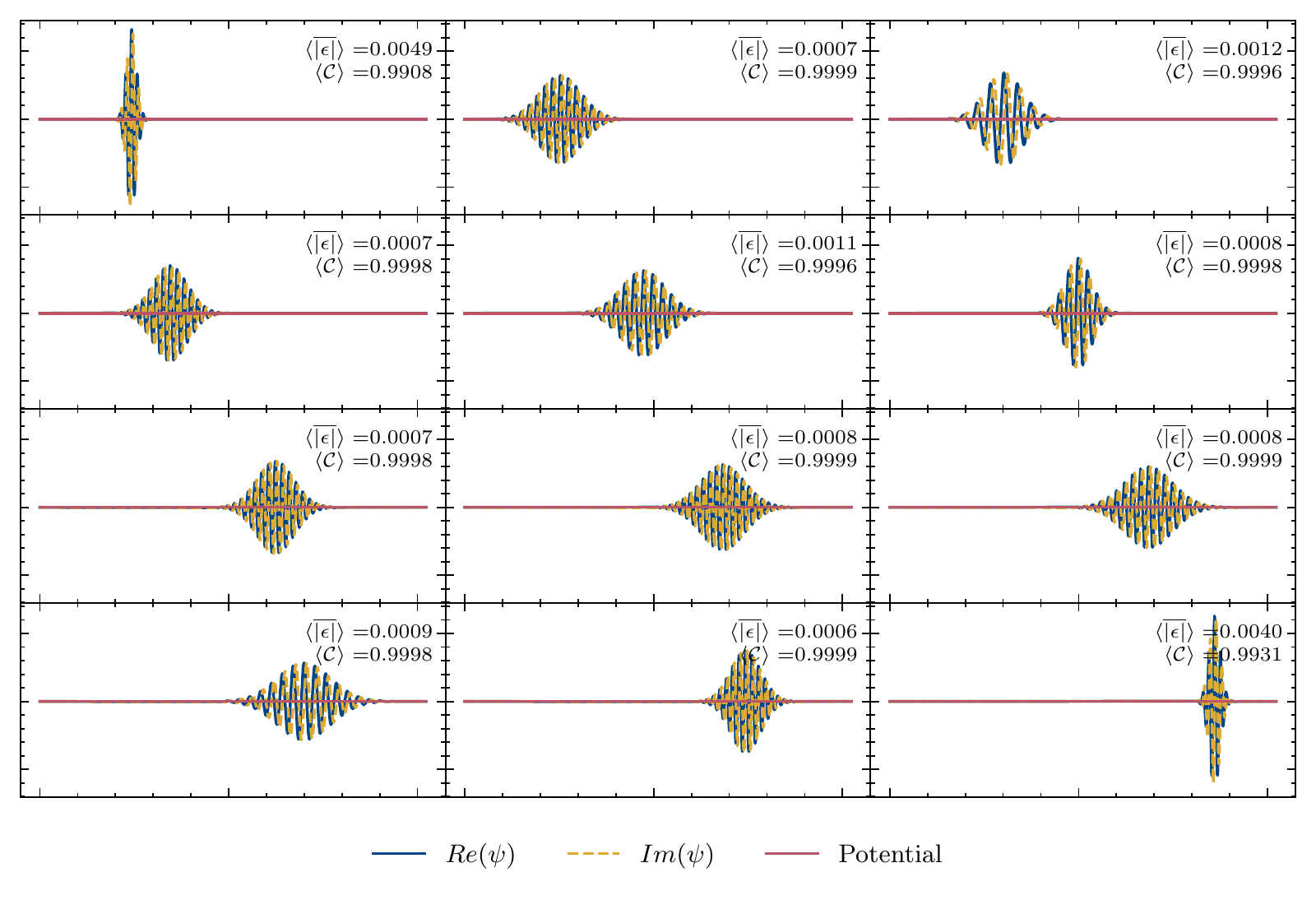}
        \end{center}
         \caption{\textbf{Freely Dispersing  Test Cases.} Fully hand-designed test data set for freely dispersing wave packets along with the performance metrics of our proposed machine learning-based emulator.} 
        \label{fig:free_tests}
        \end{figure*}

    \paragraph*{Freely Dispersing Test Cases.}
    To test the performance of our emulator in the freely dispersing regime, we have generated $12$ random test cases. The parameters were randomly selected from $X_0 \in (10.0, 90.0)$, $S_0 \in (0.5, 9.0)$, and $E_0 \in (0.0, 9.0)$. The particular test cases used in this study are depicted in Fig.~\ref{fig:free_tests}.

\begin{figure*}
        \begin{center}
            \includegraphics[width=\textwidth]{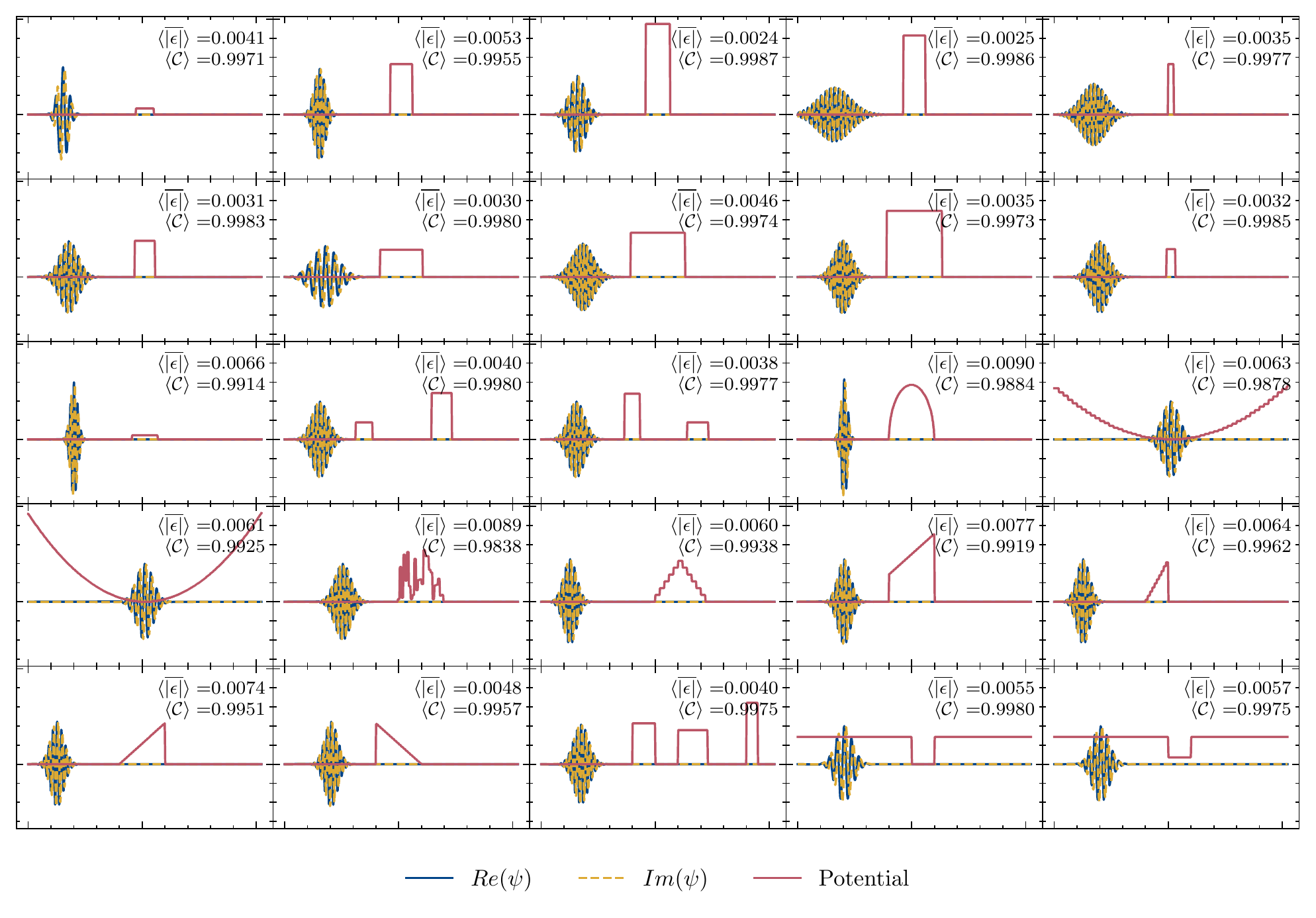}
        \end{center}
         \caption{\textbf{Real Potential Test Cases.} Full hand-designed test data set for wave packet evolution, when a real potential landscape is concerned, along with the performance metrics of our proposed machine learning-based emulator.} 
        \label{fig:barrier_tests}
        \end{figure*}

    \paragraph*{Test Cases with Potential Barriers.}
     To test the performance in the more general case, we hand-designed a test data set, that includes: 11 random single-rectangular barriers, 2 double-rectangular barriers, 1 triple-rectangular barrier, 7 irregular barriers, 2 quadratic potentials, and 2 rectangular wells. All the test cases are depicted in Fig.~\ref{fig:barrier_tests}. 
    
\cleardoublepage

\section*{Non-Gaussian Wave Packets}

    \begin{figure*}[p]
    \begin{center}
        \includegraphics[width=0.5\textwidth]{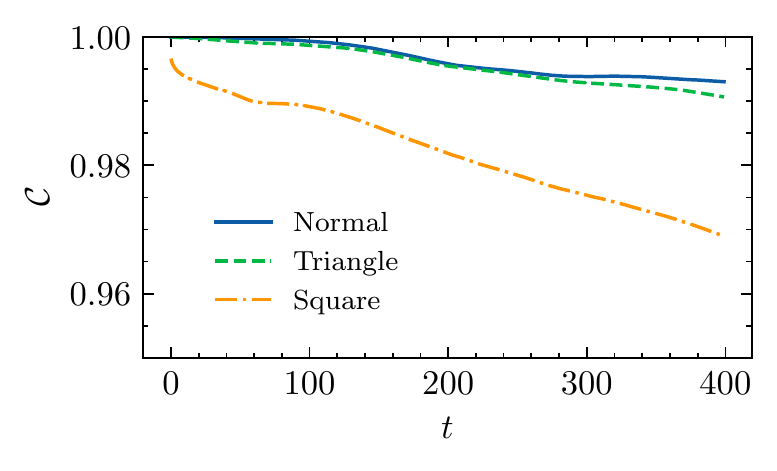}
    \end{center}
     \caption[Average direct gradients.]{{\bf Non-Gaussian Wave Packages Test.} Time-step-by-time-step error propagation for quantum wave packets with different initial modulations. Note, that the machine learning-emulator was trained only on examples depicting Gaussian wave packet evolution. To calculate the normalized correlation $\mathcal{C}$ (closer to $1$ is better), we averaged the partial results over all spatial grid points, and all our test cases including potential barriers.} 
    \label{fig:non-gaus}
    \end{figure*}

    In Figure~\ref{fig:non-gaus}, we show that the emulator can handle not only different potential landscapes, it can also generalize to different modulations of the emulated wave packet. We can see, that the performance for the triangle-shaped packet is in pair with the Gaussian-modulated packet. The performance for the square-shaped packet is noticeable worse. This can be easily understood when comparing Fourier analysis of differently shaped signals. To accurately reproduce a square signal, we need to include a number or higher harmonics. We know (cf.\ Fig.~\ref{fig:generalization} in the main text), that when energy is too high, the performance of the emulator deteriorate. In comparison, the higher harmonics (that correspond to higher energies of plain waves) are less crucial when reproducing a triangle signal with the same average accuracy as in the case of a rectangular-shaped signal. This might explains the observed differences in Fig.~\ref{fig:non-gaus}.
    
\section*{Freely Dispersing Wave Packets Case}

    In this sections, we present the performance benchmark for our neural network emulator in the case of freely dispersing quantum wave packets. As there is no potential, the input data window only contains two channels, representing the real and imaginary parts of the wave function. For the freely-dispersing-wave-packet example, we use a training data set generated by evolving 189 different Gaussian wave packets, each with a different values of $X_0$, $S_0$ and $E_0$. The machine learning models was trained for three epochs using the AdamW optimizer\cite{loshchilov2019decoupled}, and with MSE loss functions.
    
\begin{figure*}
        \begin{center}
            \includegraphics[width=\textwidth]{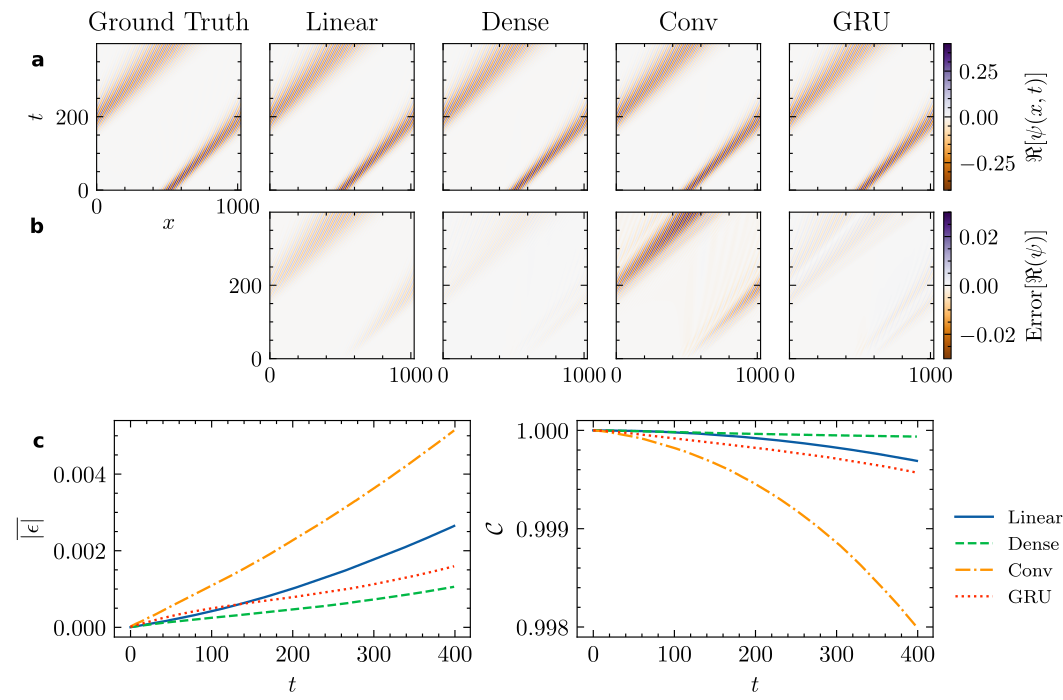}
        \end{center}
         \caption{\textbf{Freely Dispersing Quantum Wave Packet in a System with Periodic Boundary Conditions.} {\bf a}, Ground truth and the predicted solution of $\Re[\psi(x,t)]$  for emulations of the freely dispersing quantum wave packet. {\bf b}, Errors between predicted solutions and ground truth (note the scale of error is $1/10$ of the scale used to represent the wave functions in panel~{\bf a}). {\bf c}, Time-step-by-time-step error propagation in terms of the mean absolute error $\overline{|\epsilon|}$, and the normalized correlations $\mathcal{C}$.} 
        \label{fig:free_one_example}
        \end{figure*}

    Figure \ref{fig:free_one_example}{\bf a} shows results of the emulation for all four tested architectures. All four models, shown in the top row, successfully reproduce the simulation results, with only minor errors. We find that the emulation errors are adding up as time evolves. Such error accumulation is inevitable given the recurent nature of the predictions. The errors from the convolutional model are the most significant, however still within the limits of acceptable performance.

    Figure \ref{fig:free_one_example}{\bf b} and \ref{fig:free_one_example}{\bf c} shows, that all tested architectures maintain low MAEs and high correlations, even after 400 time steps, indicating that they achieve accurate and stable performance. We report the MAE and normalized correlations, averaged over all time steps and test samples in Table \ref{tab:free}.
    
    It is noteworthy that the baseline linear model achieves nearly perfect predictive powers -- as long as the potential barriers are not considered (this would be not the case if the potential barrier was present, see the results in the main text, e.g., the performance comparison presented in Fig.\ \ref{fig:barrier_one_example}). The reason is that the time evolution of the wave function in a small interval $\Delta t$, can be expressed as $\psi(t+\Delta t) = \exp \left(-i T_{x} \Delta t\right)  \psi(t) +O\left([\Delta t]^{3}\right)$, where the kinetic operator $T_{x}$ is represented by a tridiagonal matrix. Therefore, with respect to the leading terms, two consecutive time steps of the wave function evolution are linearly connected, and hence, a simple linear architecture can be trained to reproduce the simulations, given a sufficiently small time-step $\Delta t$.
   
    \begin{table*}
    \centering
    \setlength{\tabcolsep}{12pt}
    \caption{\textbf{Benchmark using Freely Dispersing Quantum Wave Packets.} Performance comparison for different architectures of our machine learning-based emulator in case of a freely dispersing quantum wave packets. As a metric, we used the mean absolute error $\overline{|\epsilon|}$ (less is better) and a normalized correlation $\mathcal{C}$ (closer to $1$ is better), both averaged over all spatial grid points, all time step, and all available test cases.}
    \begin{tabular}{llll}
    \hline \hline
    Model  & Parameters &  $\langle\overline{|\epsilon|}\rangle$  & $\langle\mathcal{C}\rangle$ \\ \hline \hline
    Linear & 2,162       &     0.0031 & 0.9827   \\ \hline
    Dense  & 12,834       &  0.0019   & 0.9952    \\ \hline
    Conv   &  19,280      &  0.0030  &  0.9966      \\ \hline
    GRU    &  19,412      &  {\bf 0.0014} &  {\bf 0.9984}        \\ \hline
    \hline
    \end{tabular}
    \label{tab:free}
    \end{table*}

\cleardoublepage

\section*{Training details and hyper-parameters tuning}
\label{sec:hyperparameters}

    All models were trained using AdamW \cite{loshchilov2019decoupled} optimizer with the mean squared error (MSE) loss function.
    We selected the initial learning rate of $10^{-3}$, with the first and second momentum $\beta_1 = 0.9$ and $\beta_2 = 0.99$, respectively. We applied a learning rate scheduler. Namely, one percent of total training steps was used as warm-up steps with a linearly increasing learning rate. Afterward, the learning rate decays linearly to $10^{-6}$.
    We provide the detailed values of the training parameters and the models hyper-parameters in Tab.~\ref{tab:training}.
    
    \begin{table*}
    \centering
    \caption{{\bf Values of training parameters and hyper-parameters.}}
    \begin{tabular}{lll}
    \hline \hline
                              & Without potential      & With potential       \\ \hline \hline
    Number of raw simulations & 189           & 2646 \\ \hline
    Input time steps (H)      & 4                  & 4                    \\ \hline
    Data window size (W)      & 23                 & 23                   \\ \hline
    Hidden size (K)               & 46                 & 69                   \\ \hline
    Training epochs           & 3                  & 5                    \\ \hline
    Optimizer                 & AdamW \cite{loshchilov2019decoupled, kingma2017adam}             & AdamW               \\ \hline
    Learning rate scheduler   & $0\rightarrow10^{-3}$ (warm-up)   & $0\rightarrow10^{-3}$ (warm-up)     \\ 
                              & $10^{-3}\rightarrow 10^{-6}$ (linear decay) & $10^{-3}
                              \rightarrow 10^{-6}$ (linear decay)   \\ \hline
    Weight decay rate         & 1                  & 1                    \\ \hline
    Gradient clipping \cite{pascanu2013difficulty}       & 1.0                  & 1.0                    \\ \hline \hline
    \end{tabular}
    \label{tab:training}
    \end{table*}

\begin{figure*}
        \begin{center}
            \includegraphics[width=\textwidth]{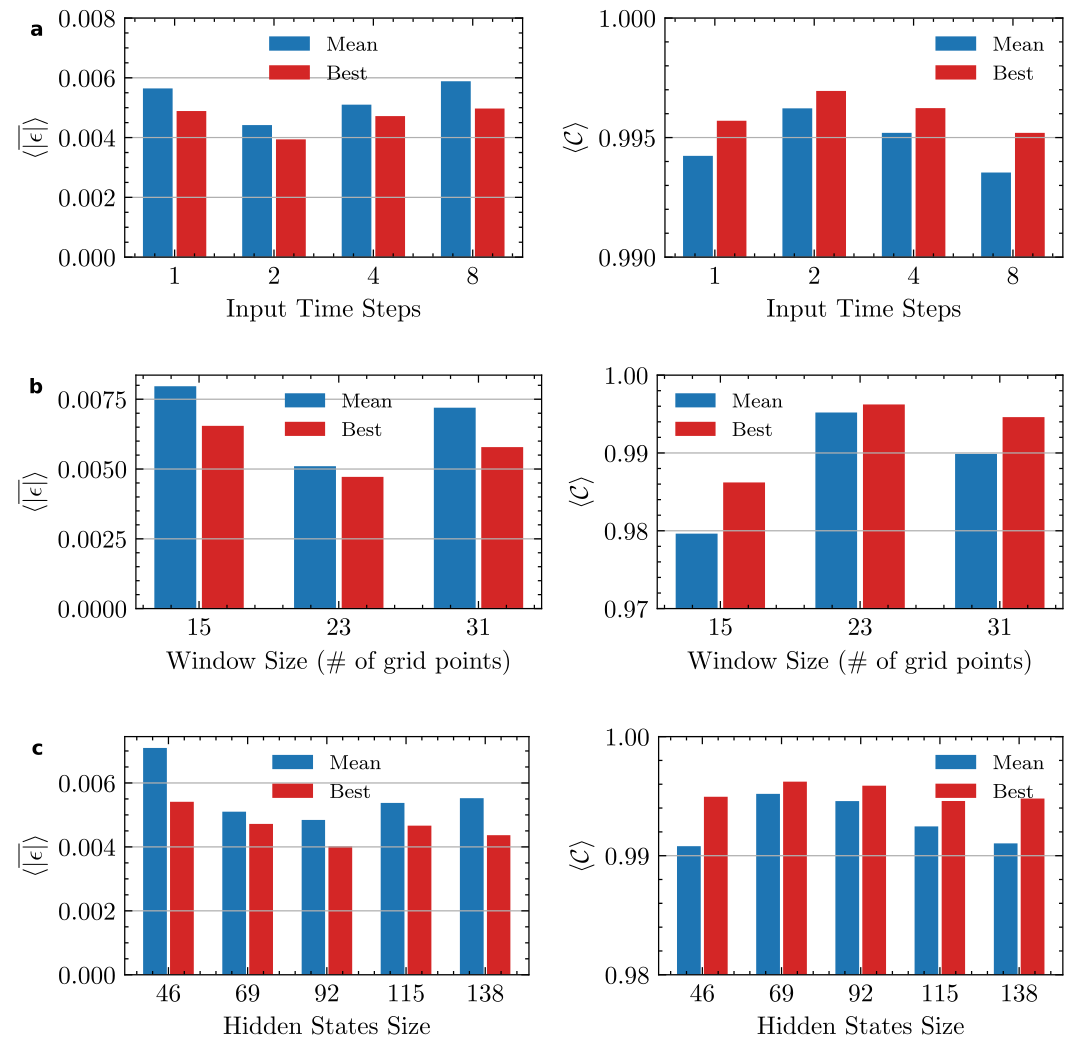}
        \end{center}
         \caption{\textbf{Hyper Parameter Tuning.} Calculated mean absolute error, $\langle\overline{|\epsilon|}\rangle$, and normalized correlation, $\langle \mathcal{C}\rangle$, averaged over test data set. {\bf a}, Number of input time steps. {\bf b}, Window size and {\bf c}, Hidden states. For each parameter, we evaluate the model with five different random seeds. The averaged results are shown in blue. The best result among five seeds is shown in red.} 
        \label{fig:ft}
        \end{figure*}

     To determine the optimal value of the hyper-parameter, we performed a search over the input time steps, the window size, and the hidden-layer size. We present the results in Fig.~\ref{fig:ft}.
    
    Figure \ref{fig:ft}{\bf a} shows that the optimal depth for the history is just two time steps, what is consistent with the analysis of the results presented in Fig.\ \ref{fig:dg}, in the main text. Noticeable, even with just one time step, our model is still able to achieve a relatively good, average performance. However, this high average score is deceiving, as it is inflated by the fact, that a single time-step is enough to predict the evolution far from the potential barrier (cf.\ results presented in Fig.\ \ref{fig:free_one_example}). However, to properly account for the scattering and tunnelling, a longer history than one time-step is needed (as evident in results presented in Fig.\ \ref{fig:barrier_one_example}, in the main text). 
    
    In Figure \ref{fig:ft}{\bf b}, we observe window size 23 gives the best results. We want to remark that this parameter is highly relevant to the sampling strategy employed during the data generation process. 

    In Figure \ref{fig:ft}{\bf c}, our model shows the lowest error with hidden states 69 and 92. The error slightly increases after hidden size 115. We expect models with a larger hidden size should in principle better approximate the target function, however, they might also be more prone to overfitting, and therefore they might require more precise optimization of the training parameters to fully leverage their potential.

\section*{Average Direct Gradients}

        \begin{figure*}
        \begin{center}
            \includegraphics[width=\textwidth]{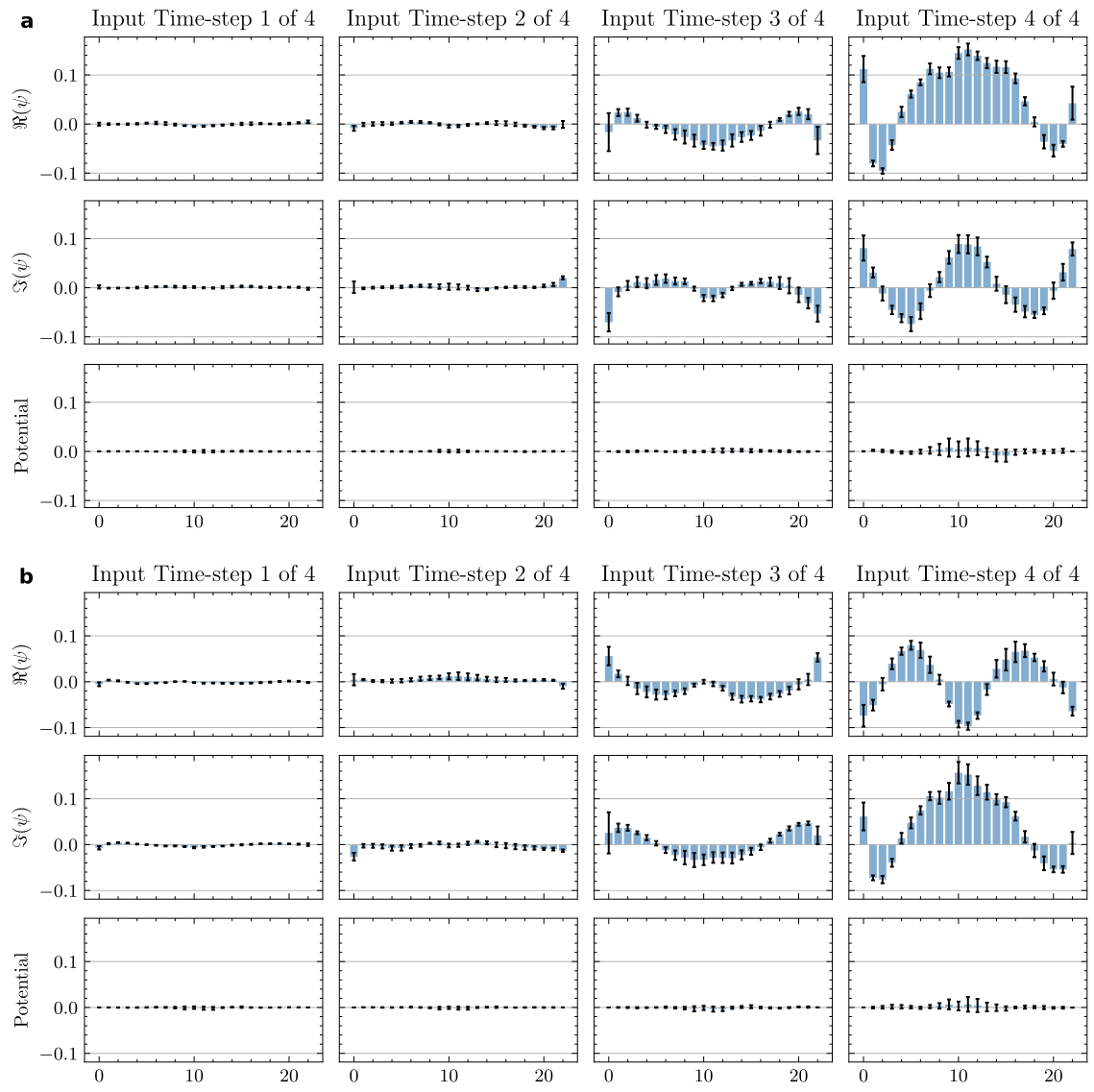}
        \end{center}
         \caption[Average direct gradients.]{{\bf Average direct gradients.} {\bf a}, The averaged direct gradients of the real part of the center pixel of the output window. {\bf b}, The averaged direct gradients of the imaginary part of the center pixel of the output window. To generate those results, we sampled 200 different data windows at different positions and time steps. The bars represent the average value, and the black line the standard deviation.} 
        \label{fig:avg_dg}
        \end{figure*}

    In Figure~\ref{fig:dg}, in the main text, we presented an exemplary results of the direct gradients measurements.
    Interestingly, when repeating the calculations for other positions of the input-window, we have consistently gotten similar values of the direct gradient. To analyze this phenomenon, we measured the average value of the direct gradients for 200 randomly sampled data windows. The results are depicted in Fig.~\ref{fig:avg_dg}. There is a clearly visible pattern, that might indicate a steady and almost linear relationship between the input wave functions and the predicted value. As presented in the previous section, with the absence of any potential barrier, the target values are linearly related to the input from the previous step. Apparently, introducing a potential does not change this linear mapping by much (compare the relatively small values of the standard deviation to the size of the bars). The situation is different when we consider the sensitivity to the values of the \emph{potential}, instead. We find that the direct gradients vary greatly over different window positions, compared to the average values. This might be an indicator, that the mapping from potential values to the target wave functions is a more complex, non-linear function -- which is also consistent with the interpretation of the results presented in Figs.\ \ref{fig:barrier_one_example} and \ref{fig:free_one_example}, as it was already discussed above.

\end{document}